\documentclass[aps,reprint,superscriptaddress,amsmath,amssymb,longbibliography]{revtex4-1}

\usepackage{graphicx}
\usepackage{amsmath}
\usepackage{bm}

\usepackage{color}
\definecolor{CLBlue}{rgb}{0, .3, .6}

\begin{document}


\title{Emergence of local irreversibility in complex interacting systems}


\author{Christopher W. Lynn}
\affiliation{Initiative for the Theoretical Sciences, Graduate Center, City University of New York, New York, NY 10016}
\affiliation{Joseph Henry Laboratories of Physics and Lewis–Sigler Institute for Integrative Genomics, Princeton University, Princeton, NJ 08544}
\author{Caroline M. Holmes}
\affiliation{Joseph Henry Laboratories of Physics and Lewis–Sigler Institute for Integrative Genomics, Princeton University, Princeton, NJ 08544}
\author{William Bialek}
\affiliation{Initiative for the Theoretical Sciences, Graduate Center, City University of New York, New York, NY 10016}
\affiliation{Joseph Henry Laboratories of Physics and Lewis–Sigler Institute for Integrative Genomics, Princeton University, Princeton, NJ 08544}
\author{David J. Schwab}
\affiliation{Initiative for the Theoretical Sciences, Graduate Center, City University of New York, New York, NY 10016}



\date{\today}

\begin{abstract}

Living systems are fundamentally irreversible, breaking detailed balance and establishing an arrow of time. But how does the evident arrow of time for a whole system arise from the interactions among its multiple elements? We show that the local evidence for the arrow of time, which is the entropy production for thermodynamic systems, can be decomposed. First, it can be split into two components: an independent term reflecting the dynamics of individual elements and an interaction term driven by the dependencies among elements. Adapting tools from non--equilibrium physics, we further decompose the interaction term into contributions from pairs of elements, triplets, and higher--order terms. We illustrate our methods on models of cellular sensing and logical computations, as well as on patterns of neural activity in the retina as it responds to visual inputs. We find that neural activity can define the arrow of time even when the visual inputs do not, and that the dominant contribution to this breaking of detailed balance comes from interactions among pairs of neurons.

\end{abstract}


\maketitle

\section{Introduction}

Living systems consume energy in order to maintain order and function. Being away from equilibrium, we expect that their microscopic dynamics violate detailed balance. Macroscopically, their behaviors define an arrow of time.  Despite recent progress in non--equilibrium statistical physics \cite{Parrondo-01, Peliti-01, Gnesotto-01}, there remain basic questions about how irreversibility at one scale emerges from collective dynamics at the scale below. To what extent does the irreversibility of a system arise from interactions between elements, rather than the independent dynamics of the elements themselves? Can simple dynamics involving pairs or triplets of elements build upon one another to generate large--scale irreversibility, thereby defining a macroscopic arrow of time, or do complex biological systems depend on higher--order combinatorial interactions?

To answer these questions, we propose a framework for decomposing the local evidence for the arrow of time in systems with many degrees of freedom. We demonstrate that the local irreversibility can be divided into two non--negative components: one that reflects the independent irreversibilities of the individual elements, and another that reflects the irreversibility due to interactions between elements. We then show that the interaction term can be further decomposed into contributions from groups of elements of different sizes, from pairs of elements to triplets to complex higher--order terms. In this way, one can determine not only whether the arrow of time arises from the dependencies between elements, but also the specific types of dynamics from which it emerges \cite{Lynn-11}. This decomposition is similar in spirit to the idea of connected correlations in the decomposition of the entropy itself \cite{Schneidman-02}.

We  apply our methods to investigate the arrow of time in neural activity. Our visual perception is built out of the patterns of electrical activity of cells in the retina, and evidence for the arrow of time must be found in these patterns. Recent experiments that record the activity of many retinal neurons simultaneously \cite{Marre-01, Palmer-01} make it possible for us to estimate all the relevant quantities directly, without introducing any model assumptions, in groups of up to five cells. We find that roughly two--thirds of these groups exhibit significant irreversibility, even when the movies shown to the retina are  completely reversible. Thus, collective neural activity can define an arrow of time even when the visual inputs do not. Moreover, across distinct stimulus ensembles, we consistently find that the local irreversibility is dominated by the dynamics of neuron pairs. Together, these results demonstrate that neuronal populations can define an arrow of time that (i) emerges primarily from pairwise dynamics and (ii) does not merely reflect the irreversibility of the stimulus.

The paper is organized as follows. In Sec. \ref{sec_S}, we define the local irreversibility and multipartite dynamics. In Sec. \ref{sec_ind_int}, we show analytically that the local irreversibility of a multipartite system can be split into two non--negative terms, the first stemming from the independent elements and the second arising from the interactions between elements. In Sec. \ref{sec_sensing}, we compare these independent and interaction irreversibilities in a simple model sensing system. In Sec. \ref{sec_dec}, we show that the irreversibility due to interactions can be further decomposed into a series of contributions from pairs of elements, triplets, and higher--order terms. In Sec. \ref{sec_logical}, we illustrate this decomposition using a minimal model of logical computations. In Sec. \ref{sec_neuron}, we apply the above methods to investigate the irreversibility of neuronal dynamics in the vertebrate retina. Finally, in Sec. \ref{sec_conclusions}, we provide conclusions and outlook, highlighting directions for future work.

\begin{figure*}
\centering
\includegraphics[width = .6\textwidth]{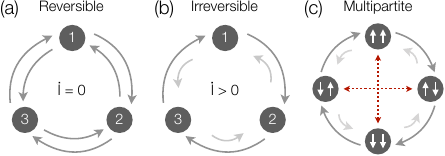} \\
\caption{Irreversibility and multipartite dynamics. (a--b) A simple three--state system, with states $x$ represented as circles and joint transition probabilities $P(x\rightarrow x')$ as arrows. (a) In a reversible system, there are no net fluxes of transitions between states, the dynamics obey detailed balance, and there is no evidence for the arrow of time. (b) Irreversible systems exhibit net fluxes between states, thereby breaking detailed balance and establishing an arrow of time. (c) A multipartite system composed of two binary spins. Only one spin is allowed to change state at a time, thus disallowing the transitions indicated by red arrows. \label{fig_multipartite}}
\end{figure*}

\section{Local irreversibility and multipartite dynamics}
\label{sec_S}

When a system is reversible, its dynamics obey detailed balance, and there are no net fluxes between states [Fig.~\ref{fig_multipartite}(a)]. By contrast, for an irreversible system, fluxes from one state to another break detailed balance [Fig.~\ref{fig_multipartite}(b)]. Critically, such irreversible dynamics establish an arrow of time: Just by observing the evolution of the system, one can distinguish whether time is flowing forward or backward.

To quantify irreversibility, consider a system with \textit{joint} transition probabilities 
\begin{equation}
P(x\rightarrow x') \equiv \text{Prob}[x_t = x,\, x_{t+1} = x'],
\end{equation}
where $x_t$ is the state of the system at time $t$. In words, this is the probability of observing the state $x$ followed by the state $x'$, and should not be confused with the \textit{conditional} transition probabilities $\text{Prob}[x_{t+1} = x'\, | \, x_t = x]$. The evidence that these dynamics carry about the arrow of time is quantified by the relative entropy, or Kullback--Leibler divergence, between the forward-- and reverse--time transition probabilities \cite{Cover-01},
\begin{equation}
\label{eq_I}
\dot{I} = \sum_{x,x'} P(x\rightarrow x') \log \left[ \frac{P(x\rightarrow x')}{P(x'\rightarrow x)} \right],
\end{equation}
where if we choose base two for the logarithms then  the evidence is measured in bits.  If a system obeys detailed balance, such that $P(x\rightarrow x') = P(x'\rightarrow x)$ for all pairs of states $x$ and $x'$, then this local irreversibility vanishes [Fig.~\ref{fig_multipartite}(a)]. Conversely, any violation of detailed balance, such that $P(x\rightarrow x') \neq P(x'\rightarrow x)$, leads to an increase in the local irreversibility [Fig.~\ref{fig_multipartite}(b)].

For Markov systems, the transition probabilities $P(x\rightarrow x')$ completely define the dynamics, and so $\dot{I}$ captures all available information about the arrow of time. Notably, if the states $x$ and $x'$ include all of the microscopic degrees of freedom in a system, then, under reasonable assumptions, Eq.~(\ref{eq_I}) defines the physical rate at which the system produces entropy \cite{Seifert-01, Ge-01}. In general, if we don't observe all the relevant degrees of freedom then the dynamics of the observable states $x$ will be non--Markovian, but  $\dot{I}$ still has a precise meaning: it represents the \textit{local} evidence for the arrow of time.

We are interested in systems where the overall state $x$ consists of states $\{x_{\rm i}\}$ for many interacting elements, ${\rm i} = 1,\, 2,\, \hdots,\, N$. Given sufficient temporal resolution, no two elements will change state at exactly the same time. In this limit, the dynamics are  defined by the joint probabilities $P(x_{\rm i}\rightarrow x_{\rm i}', \, x_{-{\rm i}})$ of one element $\rm i$ transitioning from $x_{\rm i}$ to $x_{\rm i}'$ and the rest of the system remaining in the same state, denoted $x_{-{\rm i}}$ [Fig.~\ref{fig_multipartite}(c)]. Such dynamics, which are referred to as multipartite, exhibit a number of useful properties \cite{Horowitz-01, Wolpert-01}. Chief among these properties is the fact that the local irreversibility simplifies to a sum over the individual elements:
\begin{align}
\dot{I} &= \sum_{x,x'} P(x\rightarrow x') \log \left[ \frac{P(x\rightarrow x')}{P(x'\rightarrow x)} \right] \\
&= \sum_x \sum_{{\rm i} = 1}^N \sum_{x_{\rm i}'} P(x_{\rm i}\rightarrow x_{\rm i}',\, x_{-{\rm i}}) \log \left[ \frac{P(x_{\rm i}\rightarrow x_{\rm i}',\, x_{-{\rm i}})}{P(x_{\rm i}' \rightarrow x_{\rm i},\, x_{-{\rm i}})} \right]\\
&= \sum_{{\rm i} = 1}^N \sum_{x_{-{\rm i}}} \sum_{x_{\rm i},x_{\rm i}'} P(x_{\rm i}\rightarrow x_{\rm i}',\, x_{-{\rm i}}) \log \left[ \frac{P(x_{\rm i}\rightarrow x_{\rm i}',\, x_{-{\rm i}})}{P(x_{\rm i}' \rightarrow x_{\rm i},\, x_{-{\rm i}})} \right]\\
&= \sum_{{\rm i} = 1}^N \dot{I}_{\rm i},
\end{align}
where 
\begin{equation}
\label{eq_Ii}
\dot{I}_{\rm i} = \sum_{x_{-{\rm i}}} \sum_{x_{\rm i},x_{\rm i}'} P(x_{\rm i}\rightarrow x_{\rm i}', \, x_{-{\rm i}}) \log \left[ \frac{P(x_{\rm i}\rightarrow x_{\rm i}', \, x_{-{\rm i}})}{P(x_{\rm i}' \rightarrow x_{\rm i}, \, x_{-{\rm i}})}\right]
\end{equation}
is the local irreversibility associated with element $\rm i$.

\section{Independent and interaction irreversibility}
\label{sec_ind_int}

We are now prepared to investigate the impact of interactions between elements on the irreversibility of a system. To begin, consider a hypothetical system in which the elements do not interact. In this case, the transitions of each element $\rm i$ are completely defined by the marginal transition probabilities 
\begin{equation}
P(x_{\rm i} \rightarrow x_{\rm i}') = \sum_{x_{-{\rm i}}} P(x_{\rm i} \rightarrow x_{\rm i}',\, x_{-{\rm i}}),
\end{equation}
and thus the \textit{independent} irreversibility of element $\rm i$ is given by
\begin{equation}
\label{eq_Iind}
\dot{I}^{\text{ind}}_{\rm i} = \sum_{x_{\rm i},x_{\rm i}'} P(x_{\rm i} \rightarrow x_{\rm i}') \log \left[\frac{P(x_{\rm i} \rightarrow x_{\rm i}')}{P(x_{\rm i}' \rightarrow x_{\rm i})}\right] .
\end{equation}
How does this independent irreversibility compare to the true irreversibility in Eq.~(\ref{eq_Ii})? To answer this question, we consider the difference $\dot{I}_{\rm i} - \dot{I}^{\text{ind}}_{\rm i}$, which reflects the local irreversibility of element $\rm i$ due to interactions with the rest of the system. Notably, we find that this difference---which we refer to as the \textit{interaction} irreversibility $\dot{I}^{\text{int}}_{\rm i}$ of element $\rm i$---is itself an average of KL divergences,
\begin{widetext}
\begin{align}
\label{eq_Iint_1}
\dot{I}^{\text{int}}_{\rm i} = \dot{I}_{\rm i} - \dot{I}^{\text{ind}}_{\rm i} &= \sum_{x_{-{\rm i}}} \sum_{x_{\rm i},x_{\rm i}'} P(x_{\rm i}\rightarrow x_{\rm i}', \, x_{-{\rm i}}) \left( \log\left[ \frac{P(x_{\rm i}\rightarrow x_{\rm i}', \, x_{-{\rm i}})}{P(x_{\rm i}' \rightarrow x_{\rm i}, \, x_{-{\rm i}})}\right] - \log \left[\frac{P(x_{\rm i}\rightarrow x_{\rm i}')}{P(x_{\rm i}' \rightarrow x_{\rm i})}\right] \right)\\
&= \sum_{x_{-{\rm i}}} \sum_{x_{\rm i},x_{\rm i}'} P(x_{\rm i}\rightarrow x_{\rm i}', \, x_{-{\rm i}}) \log \left[\frac{P(x_{-{\rm i}}\,| \, x_{\rm i} \rightarrow x_{\rm i}')}{P(x_{-{\rm i}}\,| \, x_{\rm i}' \rightarrow x_{\rm i})}\right] \\
\label{eq_Iint_3}
&= \sum_{x_{\rm i},x_{\rm i}'} P(x_{\rm i}\rightarrow x_{\rm i}') D_{KL} \left[P(x_{-{\rm i}}\,| \, x_{\rm i} \rightarrow x_{\rm i}')\, || \, P(x_{-{\rm i}}\,| \, x_{\rm i}' \rightarrow x_{\rm i})\right],
\end{align}
\end{widetext}
where $P(x_{-{\rm i}}\, |\, x_{\rm i} \rightarrow x_{\rm i}') = P(x_{\rm i}\rightarrow x_{\rm i}', \, x_{-{\rm i}})/P(x_{\rm i}\rightarrow x_{\rm i}')$ is the conditional probability of the state $x_{-{\rm i}}$ of the rest of the system given a transition $x_{\rm i}\rightarrow x_{\rm i}'$ in element $\rm i$.

Equation (\ref{eq_Iint_3}) immediately tells us that $\dot{I}^{\text{int}}_{\rm i} \ge 0$, thereby establishing that the presence of interactions can only increase the local irreversibility of a system. Moreover, the interaction irreversibility $\dot{I}^{\text{int}}_{\rm i}$ of element $\rm i$ admits an insightful information--theoretic interpretation: it is the amount of information that one gains about the state $x_{-{\rm i}}$ of the rest of the system by observing the forward--time dynamics of element $\rm i$ rather than the reverse--time dynamics \cite{Cover-01}. Thus, if $\rm i$'s forward-- and reverse--time dynamics contain the same information about the rest of the system, then interactions with element $\rm i$ do not contribute to the arrow of time ($\dot{I}^{\text{int}}_{\rm i} = 0$), and all of $\rm i$'s local irreversibility arises from independent dynamics ($\dot{I}_{\rm i} = \dot{I}^{\text{ind}}_{\rm i}$). Importantly, we note that Eqs.~(\ref{eq_Iint_1}--\ref{eq_Iint_3}) require multipartite dynamics; if multiple elements can change state at once, then the interaction irreversibility $\dot{I}^{\text{int}}$ is ill--defined (see Appendix \ref{app_multi}).

Together, Eqs.~(\ref{eq_Iind}--\ref{eq_Iint_3}) establish our first result: that the local irreversibility of a system can be split into two \textit{non--negative} components,
\begin{equation}
\label{eq_Iii}
\dot{I} = \dot{I}^{\text{ind}} + \dot{I}^{\text{int}},
\end{equation}
where $\dot{I}^{\text{ind}} = \sum_{{\rm i}=1}^N \dot{I}^{\text{ind}}_{\rm i}$ is the independent irreversibility of the system (reflecting the local irreversibilities of the individual elements) and $\dot{I}^{\text{int}} = \sum_{{\rm i}=1}^N \dot{I}^{\text{int}}_{\rm i}$ is the interaction irreversibility (reflecting the local irreversibility due to the dependencies between elements).

\section{Decomposing irreversibility in a sensing system}
\label{sec_sensing}

\begin{figure*}
\centering
\includegraphics[width = .9\textwidth]{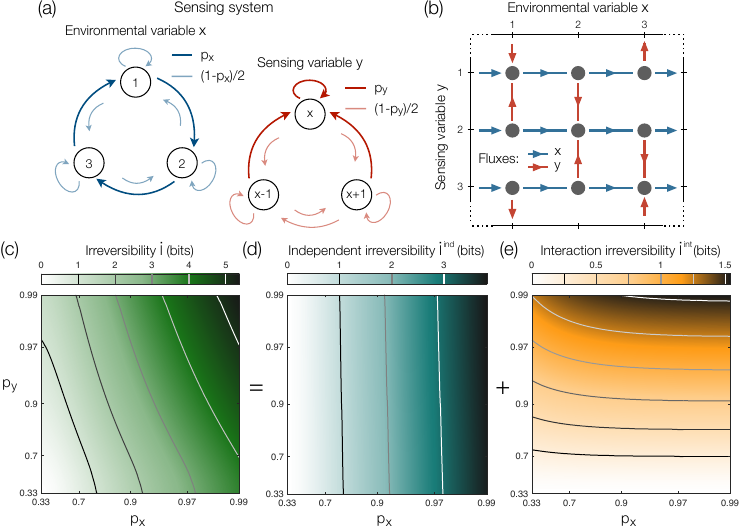} \\
\caption{Independent and interaction irreversibility in a sensing system. (a) Sensing system with an environmental variable $x$ (left) and a sensing variable $y$ (right), each with three states. At each point in time, one of the two variables is updated at random. With probability $p_x$, the environmental variable $x$ rotates clockwise, and with probability $p_y$, the sensing variable $y$ copies $x$. (b) Fluxes between states of the sensing system (for $p_x, p_y > 1/3$) induced by the environmental variable (blue) or the sensing variable (red). (c--d) Irreversibility $\dot{I}$ (c), independent irreversibility $\dot{I}^{\text{ind}}$ (d), and interaction irreversibility $\dot{I}^{\text{int}}$ (e) as functions of $p_x$ and $p_y$. While $\dot{I}$ grows with both $p_x$ and $p_y$, $\dot{I}^{\text{ind}}$ only increases with $p_x$, and $\dot{I}^{\text{int}}$ mostly increases with $p_y$, thereby distinguishing the two sources of irreversibility in the system. \label{fig_sensing}}
\end{figure*}

To illustrate the decomposition in Eq.~(\ref{eq_Iii}), we examine a sensing system, wherein a sensing variable $y$ attempts to copy an environmental variable $x$ [Fig.~\ref{fig_sensing}(a)]. Such sensing networks have been a topic of significant focus in non--equilibrium statistical mechanics \cite{Gnesotto-01, Lan-01, Barato-01, Horowitz-01, Ngampruetikorn-01}, revealing the thermodynamic costs of simple computations in living systems \cite{Gnesotto-01, Lan-01, Ngampruetikorn-01, Still-01, Still-02, Marzen-01}.

Here, we consider an environmental variable $x$ with three states and dynamics defined by
\begin{equation}
\label{eq_Px}
\hspace{-7pt} \begin{array}{cc} \hspace{48pt} x' & \\ \vspace{-9pt} & \\
P(x'\, |\, x) = \left(\begin{array}{ccc} \frac{1}{2}(1-p_x) & p_x & \frac{1}{2}(1-p_x) \\ & \vspace{-9pt} & \\ \frac{1}{2}(1-p_x) & \frac{1}{2}(1-p_x) & p_x \\ & \vspace{-9pt} & \\ p_x & \frac{1}{2}(1-p_x) & \frac{1}{2}(1-p_x) \end{array}\right) & x, \end{array}
\end{equation}
where $p_x$ is the probability of $x$ increasing from one state to the next [Fig.~\ref{fig_sensing}(a), left]. Meanwhile, the dynamics of the sensing variable $y$ are given by
\begin{equation}
\label{eq_Py}
\hspace{-6pt} \begin{array}{cc} \hspace{48pt} y' & \\ \vspace{-9pt} & \\
P(y'\, |\, x) = \left(\begin{array}{ccc} p_y & \frac{1}{2}(1-p_y) & \frac{1}{2}(1-p_y) \\ & \vspace{-9pt} & \\ \frac{1}{2}(1-p_y) & p_y & \frac{1}{2}(1-p_y) \\ & \vspace{-9pt} & \\ \frac{1}{2}(1-p_y) & \frac{1}{2}(1-p_y) & p_y \end{array}\right) & x, \end{array}
\end{equation}
where $p_y$ is the probability that $y$ copies $x$ [Fig.~(\ref{fig_sensing}(a), right]. Randomly picking one variable to update at a time, one can solve for the joint transition probabilities of the combined system $P(x,y \rightarrow x',y')$; for details see Appendix \ref{app_sensing}. Notably, since the dynamics are Markovian, $\dot{I}$ reflects the full (not just local) irreversibility of the system.

If $p_x = p_y = 1/3$, then both variables behave randomly, and the system obeys detailed balance. By contrast, if $p_x$ or $p_y > 1/3$, then the tendencies for $x$ to increase and $y$ to copy $x$ give rise to fluxes between the states of the system [Fig.~\ref{fig_sensing}(b)], thereby breaking detailed balance. Indeed, the irreversibility $\dot{I}$ increases with both $p_x$ and $p_y$ [Fig.~\ref{fig_sensing}(c)]. The independent irreversibility $\dot{I}^{\text{ind}}$, however, only increases with $p_x$, capturing the quickening dynamics of $x$ [Fig.~\ref{fig_sensing}(d)]. Meanwhile, the interaction irreversibility $\dot{I}^{\text{int}}$ primarily increases with $p_y$, capturing the strengthening dependence of $y$ on $x$ [Fig.~\ref{fig_sensing}(e)]. We therefore find that the independent irreversibility is generated by the individual motion of the environmental variable, while the interaction irreversibility arises predominantly from the dependence of the sensing variable on the environment. In this way, the decomposition in Eq.~(\ref{eq_Iii}) reveals the distinct ways that the environmental and sensing variables generate irreversibility.

\section{Irreversibility due to $k^{\text{th}}$--order dynamics}
\label{sec_dec}

Can we tell whether the arrow of time emerges from the dynamics of two or three elements at a time, or whether we require higher--order information about the system as a whole?  Answering this question requires further decomposing the local irreversibility into contributions from pairs of elements, triplets, and so on. For now, consider the marginal dynamics of pairs of elements $\rm i$ and $\rm j$; namely, the marginal transition probabilities 
\begin{eqnarray}
P(x_{\rm i} \rightarrow x_{\rm i}',\, x_{\rm j}) &=& \sum_{x_{-\{{\rm i},{\rm j}\}}} P(x_{\rm i} \rightarrow x_{\rm i}',\, x_{-{\rm i}})\\
P(x_{\rm j} \rightarrow x_{\rm j}',\, x_{\rm i}) &=& \sum_{x_{-\{{\rm i},{\rm j}\}}} P(x_{\rm j} \rightarrow x_{\rm j}',\, x_{-{\rm j}}) .
\end{eqnarray}
Imagine a hypothetical system that matches these marginal dynamics for all pairs $\rm i$ and $\rm j$, but otherwise contains minimal information about the arrow of time, so that the dynamics are maximally reversible. This minimal irreversibility, which we denote $\dot{I}^{(2)}$, sets a lower bound on the true local irreversibility $\dot{I}$, capturing all of the local irreversibility in pairs of elements and nothing more. In this way, by casting our decomposition as an optimization problem, we are able to directly translate knowledge about a system into a lower bound on its irreversibility. From a practical perspective, the local irreversibility $\dot{I}$ is convex (see Appendix \ref{app_convex}), and so there exist efficient algorithms for computing global minima. In fact, the equivalent problem of minimizing entropy production has garnered significant attention in non--equilibrium physics \cite{Schnakenberg-01, Wolpert-01, Skinner-01, Still-01}, dating back to the foundational work of Onsager and Prigogine \cite{Onsager-01, Prigogine-01}.

In general, one can compute the minimum irreversibility $\dot{I}^{(k)}$ consistent with the dynamics of $k$ elements at a time. Since these $k^{\text{th}}$--order dynamics contain all of the information about smaller groups of size $1,2,\hdots,k-1$, the minimum irreversibilities $\dot{I}^{(k)}$ form a hierarchy of lower bounds that increase toward the true local irreversiblity $\dot{I}$:
\begin{equation}
\label{eq_hier}
0 \le \dot{I}^{(1)} \le \dot{I}^{(2)} \le \hdots \le \dot{I}^{(N-1)} \le \dot{I}^{(N)} = \dot{I},
\end{equation}
where $N$ is the size of the system. There are several things to note about these inequalities. First, for thermodynamic systems, the zeroth--order bound ($\dot{I} \ge 0$) is the second law of thermodynamics, which follows from the fact that $\dot{I}$ is a KL divergence without any knowledge of the system dynamics. Second, as one might suspect, the first--order irreversibility $\dot{I}^{(1)}$---that is, the minimum irreversibility consistent with individual dynamics---is equivalent to the independent irreversibility $\dot{I}^{\text{ind}}$ (see Appendix \ref{app_Sind}). Finally, since the $N^{\text{th}}$--order dynamics contain a full description of the transition probabilities $P(x_{\rm i} \rightarrow x_{\rm i}',\, x_{-{\rm i}})$, we have $\dot{I}^{(N)} = \dot{I}$.

Inspecting the hierarchy in Eq.~(\ref{eq_hier}), we see that the local irreversibility due to $k^{\text{th}}$--order dynamics alone can be captured captured by the difference $\dot{I}^{(k)}_{\text{int}} = \dot{I}^{(k)} - \dot{I}^{(k-1)} \ge 0$, which we refer to as the interaction irreversibility of order $k$. Indeed, combining these contributions from $\dot{I}^{(1)}_{\text{int}} = \dot{I}^{(1)} = \dot{I}^{\text{ind}}$ to $\dot{I}^{(N)}_{\text{int}}$, we arrive at a full decomposition of the local irreversibility:
\begin{align}
\label{eq_dec}
\nonumber \\[-30pt]
\dot{I} = \begin{array}{c} \\ \vspace{-4pt} \\ \underbrace{\begin{array}{c} \dot{I}^{(1)}_{\text{int}} \\ \vspace{-9pt} \\ \end{array}}_{\text{\normalsize $\dot{I}^{\text{ind}}$}} \end{array} + \begin{array}{c}  \\ \vspace{-4pt} \\ \underbrace{\begin{array}{c} \dot{I}^{(2)}_{\text{int}} + \dot{I}^{(3)}_{\text{int}} + \hdots + \dot{I}^{(N)}_{\text{int}} \\ \vspace{-9pt} \\ \end{array}}_{\text{\normalsize $\dot{I}^{\text{int}}$}} \end{array},
\end{align}
which is our main contribution. We note that this decomposition is in many ways similar to the decomposition of the entropy itself into connected components \cite{Schneidman-02}.

\section{Decomposing irreversibility of logical functions}
\label{sec_logical}

To illustrate how irreversibility arises from dynamics of different orders, we apply the decomposition in Eq.~(\ref{eq_dec}) to a class of noisy logical functions. Specifically, we consider binary variables $x$ and $y$ that change state at each time step with probability $p_{\text{flip}}$, and third binary variable  $z$ that is the output of a logical function with a probability of error $p_{\text{error}}$ [Fig.~\ref{fig_logical_1}(a); see Appendix \ref{app_logical} for a full description]. As for the sensing system in Fig.~\ref{fig_sensing}, because the dynamics are Markovian the local irreversibility $\dot{I}$ represents the full irreversibility of the system.

We note that binary variables in steady state, such as those considered here, cannot break detailed balance on their own (Appendix \ref{app_bin}). Thus, for binary steady--state systems, the independent irreversibility vanishes ($\dot{I}^{\text{ind}}=0$), such that the arrow of time arises entirely from the interactions between the elements, $\dot{I} = \dot{I}^{\text{int}} = \dot{I}^{(2)}_{\text{int}} + \hdots + \dot{I}^{(N)}_{\text{int}}$. Specifically for the logical functions [Fig.~\ref{fig_logical_1}(a)], there are only two contributions to the irreversibility: that due to pairwise dynamics $\dot{I}^{(2)}_{\text{int}}$ and that due to the full triplet dynamics $\dot{I}^{(3)}_{\text{int}}$.

\begin{figure}
\centering
\includegraphics[width = .9\columnwidth]{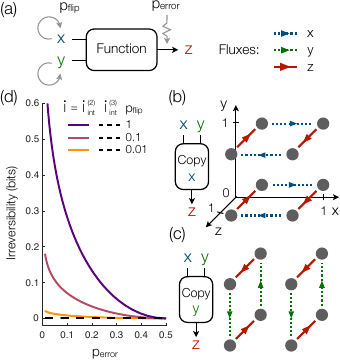} \\
\caption{Decomposing irreversibility in noisy logical functions. (a) System of three binary variables $x$, $y$, and $z$, where $z$ performs a noisy logical function on the inputs $x$ and $y$. At each point in time, one of the variables is updated at random. With probability $p_{\text{flip}}$, the inputs $x$ and $y$ change value, and with probability $p_{\text{error}}$, the output $z$ fails to perform the specified function (see Appendix \ref{app_logical}). (b--c) Fluxes between states of the system when $z$ either copies $x$ (b) or copies $y$ (c). (d) Irreversibilities $\dot{I}$ (full), $\dot{I}^{(2)}_{\text{int}}$ (pairwise interaction), and $\dot{I}^{(3)}_{\text{int}}$ (triplet interaction) versus $p_{\text{error}}$ for different values of $p_{\text{flip}}$. For both of the copy functions, irreversibility arises entirely from pairwise dynamics ($\dot{I} = \dot{I}^{(2)}_{\text{int}}$), while triplet dynamics do not contribute ($\dot{I}^{(3)}_{\text{int}} = 0$). \label{fig_logical_1}}
\end{figure}

To begin, consider a simple function where $z$ copies either $x$ or $y$ while ignoring the other input [Fig.~\ref{fig_logical_1}(b--c)]; these are binary simplifications of the sensing system in Fig.~\ref{fig_sensing}. As $p_{\text{error}}$ increases---that is, as the accuracy of the function decreases---we find that the irreversibility $\dot{I}$ decreases [Fig.~\ref{fig_logical_1}(d)]. Indeed, as $p_{\text{error}}$ approaches $1/2$, the output $z$ completely decouples from the inputs $x$ and $y$, and the system becomes reversible ($\dot{I} = 0$). Additionally, the arrow of time vanishes if the inputs $x$ and $y$ are static ($p_{\text{flip}} = 0$) and grows as the inputs become more dynamic [that is, as $p_{\text{flip}}$ increases; Fig.~\ref{fig_logical_1}(d)]. Visualizing the fluxes between states of the system, we see that the tendency for $z$ to copy $x$ (equivalently, $y$) only induces fluxes in the $x$--$z$ (or $y$--$z$) plane [Fig.~\ref{fig_logical_1}(b--c)]. Accordingly, for all values of $p_{\text{flip}}$ and $p_{\text{error}}$, the irreversibility arises entirely from pairwise dynamics ($\dot{I} = \dot{I}^{(2)}_{\text{int}}$), while triplet dynamics do not contribute to the irreversibility [$\dot{I}^{(3)}_{\text{int}} = 0$; Fig.~\ref{fig_logical_1}(d)].

\begin{figure}
\centering
\includegraphics[width = .9\columnwidth]{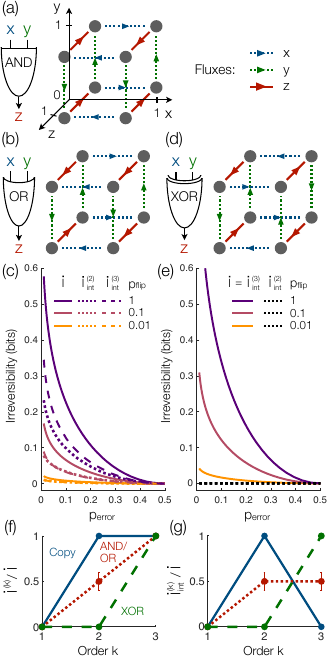} \\
\caption{Decomposing irreversibility in AND, OR, and XOR functions. (a--b) For logical systems defined as in Fig.~\ref{fig_logical_1}(a), we illustrate the fluxes between states when $z$ executes either AND (a) or OR (b). (c) Irreversibilities $\dot{I}$ (full), $\dot{I}^{(2)}_{\text{int}}$ (pairwise interaction), and $\dot{I}^{(3)}_{\text{int}}$ (triplet interaction) of the AND and OR systems versus $p_{\text{error}}$ for different values of $p_{\text{flip}}$. Irreversibility arises from a combination of pairwise ($\dot{I}^{(2)}_{\text{int}} > 0$) and triplet ($\dot{I}^{(3)}_{\text{int}} > 0$) dynamics. (d--e) Fluxes (d) and irreversibilities (e) when $z$ performs XOR. Irreversibility arises entirely from triplet dynamics ($\dot{I} = \dot{I}^{(3)}_{\text{int}}$), while pairwise dynamics are reversible ($\dot{I}^{(2)}_{\text{int}} = 0$). (f--g) Minimum irreversibilities $\dot{I}^{(k)}$ (f) and interaction irreversibilities $\dot{I}^{(k)}_{\text{int}}$ (g), normalized by the full irreversibility $\dot{I}$, as functions of the order $k$ for different logical functions. The errorbars for AND and OR reflect the small variability in $\dot{I}^{(k)}/\dot{I}$ and $\dot{I}^{(k)}_\text{int}/\dot{I}$ over the range of different $p_\text{error}$ and $p_\text{flip}$ values. \label{fig_logical_2}}
\end{figure}

For comparison, consider the AND and OR functions [Fig.~\ref{fig_logical_2}(a--b)]. Just as for the copy functions (Fig.~\ref{fig_logical_1}), the irreversibilities of AND and OR (which we note are identical) increase both with the accuracy of the system (as $p_{\text{error}}$ decreases) and with the speed of dynamics [as $p_{\text{flip}}$ increases; Fig.~\ref{fig_logical_2}(c)]. However, in contrast to the copy functions, the full dynamics of AND and OR cannot be deduced from pairs of variables alone. Thus, the irreversibility arises from a combination of both pairwise and triplet dynamics [Fig.~\ref{fig_logical_2}(c)]. Finally, for the XOR function, the behavior of the system only becomes apparent when all three variables are observed simultaneously [Fig.~\ref{fig_logical_2}(d)]. As such, the irreversibility of XOR arises entirely from triplet dynamics ($\dot{I} = \dot{I}^{(3)}_{\text{int}}$), while the pairwise dynamics are completely reversible [$\dot{I}^{(2)}_{\text{int}} = 0$; Fig.~\ref{fig_logical_2}(e)].  This is consistent with the status of XOR as the prototype of combinatorial interactions.

The results of this section are summarized in Fig.~\ref{fig_logical_2}(f--g), where we plot the minimum irreversibilities $\dot{I}^{(k)}$ [Fig.~\ref{fig_logical_2}(f)] and interaction irreversibilities $\dot{I}^{(k)}_{\text{int}}$ [Fig.~\ref{fig_logical_2}(g)] of the different logical functions, normalized by the full irreversibilities $\dot{I}$. Since the systems all consist of binary steady--state dynamics, the first--order irreversibilities $\dot{I}^{(1)} = \dot{I}^{(1)}_{\text{int}}$ vanish, and therefore the independent dynamics do not define an arrow of time ($\dot{I}^{\text{ind}} = 0$). For the copy functions [Fig.~\ref{fig_logical_1}(b--c)], irreversibility is driven entirely by second--order dynamics; for the AND and OR functions [Fig.~\ref{fig_logical_2}(a--b)], the arrow of time arises from a combination of second-- and third--order dynamics; and for the XOR function [Fig.~\ref{fig_logical_2}(d)], irreversibility is driven entirely by third--order dynamics [see Fig.~\ref{fig_logical_2}(g)]. In this way, the decomposition in Eq.~(\ref{eq_dec}) can be used to uncover the order of the dynamics that generate irreversibility in interacting systems.

\section{Decomposing irreversibility in neuronal populations}
\label{sec_neuron}

Using the framework developed above, we are ultimately interested in understanding how irreversibility emerges in biological systems. Here, we study electrical activity in groups of neurons at the output of the retina. These ganglion cells provide all the data that the brain has about the visual world, and hence their state provides the ingredients out of which visual perceptions are synthesized, including our perception of the arrow of time. Importantly, information about visual stimuli is encoded not just in the firing of individual neurons, but also in the web of dependencies between neurons \cite{Strong-01, Schneidman-01, Palmer-01}. It remains unknown, however, whether groups of neurons exhibit fluxes between collective states---thereby breaking detailed balance---and if so, whether such irreversibility arises from pairs of neurons or from complicated higher--order dynamics.

\begin{figure}
\centering
\includegraphics[width = \columnwidth]{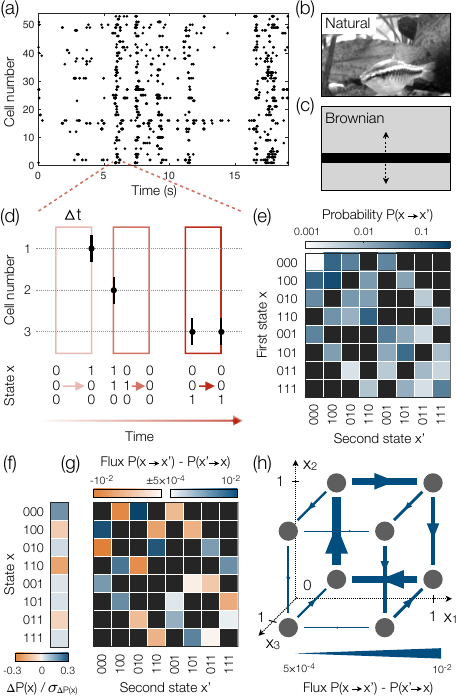} \\
\caption{Broken detailed balance in a group of neurons. (a) Dots mark the times of action potentials (``spikes'') from 53 neurons in the salamander retina responding to a visual stimulus (see Appendix \ref{app_neuron} for experimental details). (b--c) The same 53 neurons are exposed to three different stimuli: a natural movie of fish swimming (b), a horizontal bar whose movement is defined by a Brownian particle on a spring (c), and the same Brownian bar in panel (c) but with one trajectory repeated multiple times. (d) To record multipartite transitions [see Fig.~\ref{fig_multipartite}(c)], we slide a window of width $\Delta t = 20\,{\rm ms}$ along the time series. A neuron transitions to active when a spike enters the front of the window (left), or inactive when a spike exits the back of the window (center). Self--transitions can occur when a cell spikes twice within the same window (right). (e--h) A random group of three neurons responding to a natural movie. (e) Probabilities $P(x\rightarrow x')$, where black entries indicate transitions that are disallowed under multipartite dynamics. (f) Changes in state probabilities $\Delta P(x)$ are small relative to their standard deviations $\sigma_{\Delta P(x)}$, indicating that the system is in steady state. (g--h) Fluxes $P(x\rightarrow x') - P(x'\rightarrow x)$ illustrated as a matrix (g) and as a directed network (h). The presence of fluxes demonstrates that the neurons break detailed balance, defining an arrow of time. \label{fig_neuron_1}}
\end{figure}

Here we analyze experiments on the salamander retina [Fig.~\ref{fig_neuron_1}(a)], where it is possible to record form many neurons simultaneously as they respond to complex visual stimuli  \cite{Marre-01}. These experiments explored three very different kinds of visual inputs: natural movies [Fig.~\ref{fig_neuron_1}(b)], a single horizontal bar whose vertical motion is equivalent to a Brownian particle on a spring [Fig.~\ref{fig_neuron_1}(c)], and the Brownian bar with precise repetitions of the same trajectory. Although this was not the goal of the original experiments, we note that the natural movies violate time--reversal invariance, being easily recognized when played forward vs. backward, while the Brownian bar is an equilibrium system and obeys detailed balance. Appendix \ref{app_neuron} gives a more detailed description of the experimental setup and procedures from Ref \cite{Marre-01}.

\subsection{Broken detailed balance in neuronal dynamics}

The problems of detecting and quantifying irreversibility in data have garnered significant attention in the statistical mechanics of living systems \cite{Gnesotto-01, Battle-01, Lynn-09, Martinez-01, Skinner-01, Li-01}. To detect irreversibility, one must simply search for violations of detailed balance; namely, fluxes between the states of a system \cite{Gnesotto-01, Battle-01, Lynn-09}. To quantify the irreversibility of a system, however, one must estimate or bound $\dot{I}$ from time--series measurements \cite{Lynn-09, Martinez-01, Skinner-01, Li-01}. Here, in addition to estimating the local irreversibility $\dot{I}$, we further wish to decompose $\dot{I}$ into contributions from dynamics of various orders [as in Eq.~(\ref{eq_dec})]. In order to do so---that is, in order to compute the minimum irreversibilities $I^{(k)}$ consistent with $k^{\text{th}}$--order dynamics---we must begin by estimating the transition probabilities $P(x_{\rm i}\rightarrow x_{\rm i}', \, x_{-{\rm i}})$ themselves.

We consider a neuron $\rm i$ active ($x_{\rm i} = 1$) if it generates an action potential (``spike'') at least once within a time window of width $\Delta t = 20\,{\rm ms}$, or inactive ($x_{\rm i} = 0$) if it is silent. In this way, the collective state of $N$ neurons is a binary vector $x = \{x_1, x_2,\hdots, x_N\}$. As we slide the window along the time series, it is almost always the case that only one cell $\rm i$ changes state at a time, either by having a spike enter the front of the window [Fig.~\ref{fig_neuron_1}(d), left window] or exit the back of the window [Fig.~\ref{fig_neuron_1}(d), center window]. Each time this occurs, we record a new transition between states $x\rightarrow x'$. For completeness, we remark that self--transitions can occur when a cell spikes twice within the same window [Fig.~\ref{fig_neuron_1}(d), right window]; however, we note that the all--silent state $\{0,\hdots,0\}$ cannot have self--transitions, since no spike enters or exits the sliding window. In the rare instances when two spikes enter or exit the window at exactly the same time (within the experimental resolution of $0.1\,{\rm ms}$), we break ties by adding small random noise to the spike times, thus yielding multipartite dynamics wherein only one cell changes state at a time.

For example, in Fig.~\ref{fig_neuron_1}(e) we illustrate the probabilities of transitions between the states of $N=3$ neurons responding to a natural movie [Fig.~\ref{fig_neuron_1}(b)]. Notably, the changes in state probabilities $\Delta P(x) = \sum_{x'} P(x'\rightarrow x) - P(x \rightarrow x')$ are small relative to errors [Fig.~\ref{fig_neuron_1}(f)], indicating that the group of neurons is in a stochastic steady state. As discussed above (and in Appendix \ref{app_bin}), binary steady--state variables cannot break detailed balance on their own. Thus, even though neurons violate detailed balance at the subcellular scale, at the coarse-grained level of binary activity the individual neurons in Fig.~\ref{fig_neuron_1}(e--f) do not define a local arrow of time. However, when examined as a group, we find that the three cells exhibit fluxes between collective states [Fig.~\ref{fig_neuron_1}(g--h)], thereby breaking detailed balance. In combination, these results establish that the group of neurons operates at a non--equilibrium steady state.

\subsection{Local irreversibility depends on stimulus}

We are now prepared to estimate the collective irreversibility of groups of neurons. We note that neurons---indeed, biological systems generally---can have long--range temporal dependencies. Thus, in contrast to the Markov systems examined in previous sections (Figs.~\ref{fig_sensing}--\ref{fig_logical_2}), here $\dot{I}$ reflects the local (rather than total) irreversibility of the system. As with other information--theoretic quantities, estimating the local irreversibility from data is challenging, and prone to systematic errors due to finite data. As described in Appendix \ref{app_finite}, we find that these can be controlled using the strategy of Ref.~\cite{Strong-01} if we restrict our attention to groups of no more than $N=5$ cells.

\begin{figure}
\centering
\includegraphics[width = \columnwidth]{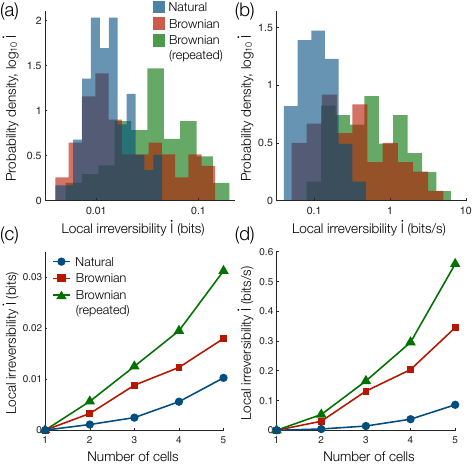} \\
\caption{Stimulus--dependence of local irreversibility. (a) Distributions of local irreversibilities $\dot{I}$  for 5--cell groups responding to a natural movie (blue), a Brownian bar (red), and a repeated Brownian bar (green). (b) The same as panel (a), but normalized to bits per second to account for variations in spike rates across stimuli. In panels (a--b), out of 100 random 5--cell groups, we only include those with significant local irreversibility (see Appendix \ref{app_finite}). (c--d) Local irreversibility (c) and after normalizing to bits per second (d) for different stimuli as functions of the number of cells in a group. Data points are averaged over 100 random groups. \label{fig_neuron_2}}
\end{figure}

\begin{figure*}
\centering
\includegraphics[width = .95\textwidth]{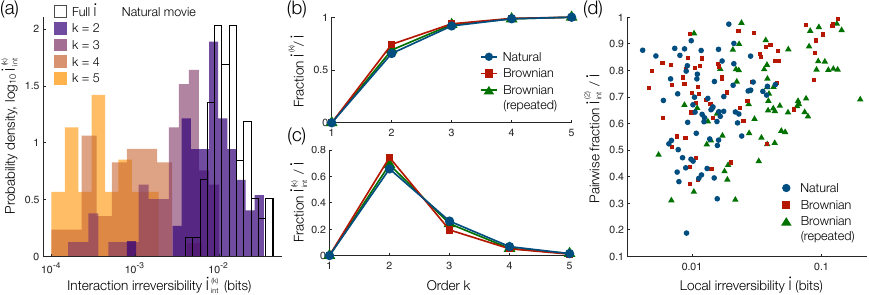} \\
\caption{Decomposing local irreversibility in neuronal activity. (a) Distributions of interaction irreversibilities $\dot{I}^{(k)}_{\text{int}}$ of different orders $k$ for 5--cell groups responding to a natural movie. (b--c) Minimum irreversibilities $\dot{I}^{(k)}$ (b) and interaction irreversibilities $\dot{I}^{(k)}_{\text{int}}$ (c), normalized by the true local irreversibilities $\dot{I}$, as functions of the order $k$ averaged over 5--cell groups. (d) The fraction of pairwise irreversibility $\dot{I}^{(2)}_{\text{int}}/\dot{I}$ increases significantly with the local irreversibility $\dot{I}$ for 5--cell groups (Spearman coefficient $r = 0.31$, $p < 10^{-3}$). In all panels, out of 100 random groups, we only include those with significant local irreversibility for each stimulus. \label{fig_neuron_3}}
\end{figure*}

After correcting for finite--data effects, out of 100 random 5--cell groups, across the different stimuli we find that $60$--$68\%$ exhibit significant local irreversibility $\dot{I}$, thereby defining an arrow of time. Moreover, for all cell groups and all stimuli, we find that the dynamics are in steady state (see Appendix \ref{app_ss}), indicating that individual cells do not break detailed balance, and therefore that any local arrow of time arises from the collective dynamics of multiple neurons. Surprisingly, despite the fact that the Brownian bar is completely reversible, neuronal dynamics are more irreversible when responding to this stimulus than the natural movie [Fig.~\ref{fig_neuron_2}(a)]. Moreover, the local irreversibility is even larger when the same Brownian trajectory is repeated multiple times [Fig.~\ref{fig_neuron_2}(a)], suggesting that a repeated input can induce a stronger arrow of time in the neuronal responses. We confirm that these differences in local irreversibility hold even after accounting for variations in the overall rate of spiking across the stimulus ensembles [Fig.~\ref{fig_neuron_2}(b)]. Additionally, the same ordering of stimuli holds for all group sizes from $N=2$ to $N=5$ cells [Fig.~\ref{fig_neuron_2}(c--d)]. These results demonstrate that the arrow of time in neuronal activity does not simply reflect the irreversibility of the stimulus. Instead, neuronal dynamics can define an arrow of time even when the stimulus does not.

\subsection{Local irreversibility arises from low--order dynamics}

To implement the decomposition of local irreversibility from Eq.~(\ref{eq_dec}), we need numerical methods to construct the probability distributions that minimize $\dot{I}$ while matching the observed $k^{\text{th}}$--order dynamics. We provide one such method for binary systems in Appendix \ref{app_min}.

To recall, for binary steady--state systems, the independent irreversibility $\dot{I}^{\text{ind}}$ vanishes, and so the local irreversibility arises entirely from the collective dynamics of two or more variables (see Appendix \ref{app_bin}). For groups of $N=5$ cells responding to the natural movie, we find that pairwise dynamics account for much more of the local irreversibility than higher--order dynamics [Fig.\ref{fig_neuron_3}(a)]. In fact, across all stimuli, pairwise dynamics generate $66$--$74\%$ of the local irreversibility [Fig.~\ref{fig_neuron_3}(b)], more than 3$^{\text{rd}}$--, 4$^{\text{th}}$--, and 5$^{\text{th}}$--order dynamics combined [Fig.~\ref{fig_neuron_3}(c)]. Moreover, the fraction of the irreversibility captured by pairwise dynamics increases significantly with the local irreversibility itself [Fig.~\ref{fig_neuron_3}(d)], demonstrating that groups of neurons that operate further from detailed balance do so in an even more pairwise fashion. 
Perhaps most notably, despite the fact that the magnitude of the local irreversibility varies significantly from one stimulus to another (Fig.~\ref{fig_neuron_2}), we find that the proportions of irreversibility captured by different types of dynamics remain consistent across stimuli [Fig.~\ref{fig_neuron_3}(b--c)].

In combination, the results of this section indicate that the arrow of time in retinal neurons (i) varies depending on the specific stimulus (Fig.~\ref{fig_neuron_2}), yet (ii) does not simply reflect the irreversibility of the stimulus, and (iii) consistently arises from the same combination of low--order dynamics, driven primarily by pairs of neurons (Fig.~\ref{fig_neuron_3}).

\section{Conclusions}
\label{sec_conclusions}

Irreversible dynamics support a wide range of biological functions, yet it remains unclear how macroscopic irreversibility arises from the microscopic dynamics of individual components. In this study, we propose a framework to uncover how irreversibility emerges in complex interacting systems. To do so, we develop analytic and numerical techniques for decomposing the information--theoretic evidence for the arrow of time into contributions from individual elements, pairs, and higher--order dynamics. We illustrate our methods on the examples of irreversible dynamics in models for sensing systems (Fig.~\ref{fig_sensing}) and logical functions (Figs.~\ref{fig_logical_1} and \ref{fig_logical_2}).  Moving to real data, we find that the irreversibility of retinal neurons varies from one stimulus to another, but consistently arises from pairwise dynamics (Figs.~\ref{fig_neuron_1}--\ref{fig_neuron_3}).

These results suggest several new directions. For example, given that the irreversibility of retinal neurons does not simply reflect that of the stimulus, it is natural to wonder which stimulus properties are, in fact, responsible for inducing irreversibility in groups of cells. Additionally, in the process of decomposing the local irreversibility, one must compute a hierarchy of minimally irreversible models consistent with observed $k^{\text{th}}$--order dynamics. Just as maximum entropy models have been successful in describing distributions over states at single moments in time \cite{Schneidman-01,Bialek-02,Meshulam-17,Lynn-04}, might these minimum irreversibility models provide insights into the dynamical flow of living systems from one state to another? More generally, we remark that the proposed framework is non--invasive, applying to any system with time--series data. Thus, the methods can be used to examine irreversible dynamics in a wide range of other biological systems, from molecular and cellular networks \cite{Ngampruetikorn-01, Lan-01, Stuhrmann-01, Soares-01, Gnesotto-01, Fakhri-01}, to large--scale recordings in the brain \cite{Lynn-09, Lynn-05}, to entire populations of animals and humans \cite{Bialek-02, Cavagna-01, Lynn-04}.

\begin{acknowledgments}
We thank SE Palmer for helpful discussions and for guiding us through the data of Ref.~\cite{Palmer-01}. This work was supported in part by the National Science Foundation, through the Center for the Physics of Biological Function (PHY--1734030) and a Graduate Research Fellowship (CMH); by the National Institutes of Health through the BRAIN initiative (R01EB026943); by the James S McDonnell Foundation through a Postdoctoral Fellowship Award (CWL); by the Simons Foundation; and by a Sloan Research Fellowship (DJS).
\end{acknowledgments}

\appendix

\section*{CITATION DIVERSITY STATEMENT}

Recent work in several fields of science \cite{Mitchell-01, Dion-01, Caplar-01, Dworkin-01, Bertolero-01}, and physics in particular \cite{Teich-01}, has identified citation bias negatively impacting women and minorities. Here we sought to proactively consider choosing references that reflect the diversity of the field in thought, form of contribution, gender, and other factors. Excluding (including) self--citations to the current authors, our references contain 26\% (26\%) women lead authors and 29\% (33\%) women senior authors.

\section{Multipartite dynamics required to decompose irreversibility}
\label{app_multi}

In Sec. \ref{sec_ind_int}, we show that the local irreversibility $\dot{I}$ of a multipartite system can be decomposed into two non--negative components [Eq.~(\ref{eq_Iii})]: the independent irreversibility $\dot{I}^{\text{ind}}$ and the interaction irreversibility $\dot{I}^{\text{int}}$. Here we show that multipartite dynamics are necessary for this decomposition. Specifically, we establish that if multiple elements are allowed to change state at the same time, then the decomposition in Eq.~(\ref{eq_Iii}) can break down and the interaction irreversibility can become ill--defined.

Consider, for example, a system with two identical elements $x$ and $y$, such that $x=y$ at all moments in time. Given the joint transition probabilities of one of the variables $P(x\rightarrow x')$, the dynamics of the combined system are given by
\begin{equation}
P(x,y\rightarrow x',y') = P(x\rightarrow x')\delta_{x,y}\delta_{x',y'}.
\end{equation}
For such a system, the irreversibility is given by
\begin{align}
\dot{I} &= \sum_{x,x',y,y'} P(x,y\rightarrow x',y') \log \frac{P(x,y\rightarrow x',y')}{P(x',y'\rightarrow x,y)} \\
&= \sum_{x,x'} P(x\rightarrow x') \log \frac{P(x\rightarrow x')}{P(x'\rightarrow x)}.
\end{align}
To compute the independent irreversibility $\dot{I}^{\text{ind}}$, we note that the marginal dynamics of $y$ are identical to that of $x$:
\begin{align}
P(y\rightarrow y') &= \sum_{x,x'} P(x,y\rightarrow x',y') \\
&= \sum_{x,x'} P(x\rightarrow x')\delta_{x,y}\delta_{x',y'} \\
&= P(x\rightarrow x').
\end{align}
Thus, the independent irreversibility is given by
\begin{align}
\dot{I}^{\text{ind}} &= \dot{I}^{\text{ind}}_x + \dot{I}^{\text{ind}}_y \\
&= \sum_{x,x'} P(x\rightarrow x') \log \frac{P(x\rightarrow x')}{P(x'\rightarrow x)} \\
& \quad\quad\quad + \sum_{y,y'} P(y\rightarrow y') \log \frac{P(y\rightarrow y')}{P(y'\rightarrow y)} \\
&= 2\sum_{x,x'} P(x\rightarrow x') \log \frac{P(x\rightarrow x')}{P(x'\rightarrow x)} \\
&= 2\dot{I}.
\end{align}
Since $\dot{I}^{\text{int}} = \dot{I} - \dot{I}^{\text{ind}} = -\dot{I}$, we find that the interaction irreversibility is negative, thus violating the decomposition of the local irreversibility into non--negative terms.

\section{Solving the sensing system}
\label{app_sensing}

Consider a sensing system composed of an environmental variable $x$ and a sensing variable $y$, each with three states. The environmental variable $x$ increases with probability $p_x$, and the sensing variable $y$ copies $x$ with probability $p_y$, yielding the dynamics in Eqs.~(\ref{eq_Px}--\ref{eq_Py}). Randomly choosing one variable to update at each point in time, the dynamics of the combined system are defined by the conditional transition probabilities
\begin{equation}
\label{eq_sensing1}
P(x',y'\, | \, x, y) = \frac{1}{2}\big( P(x'\, |\, x)\delta_{y,y'} + P(y'\, |\, x)\delta_{x,x'}\big).
\end{equation}
Using the stationary condition $\pi(x,y) = \sum_{x',y'}P(x,y\, |\, x', y')\pi(x',y')$, one can solve for the stationary distribution:
\begin{equation}
\label{eq_sensing2}
\hspace{-5pt} \pi(x,y) \propto \left(\begin{array}{c} 2 + 7p_y + p_x(1 + 3p_y + 3p_x) \\ 5 - 2p_y + p_x(4 - 6p_y + 3p_x) \\ 6 - 5p_y + p_x(1 + 3p_y + 3p_x) \\ 6 - 5p_y + p_x(1 + 3p_y + 3p_x) \\ 2 + 7p_y + p_x(1 + 3p_y + 3p_x) \\ 5 - 2p_y + p_x(4 - 6p_y + 3p_x) \\ 5 - 2p_y + p_x(4 - 6p_y + 3p_x) \\ 6 - 5p_y + p_x(1 + 3p_y + 3p_x) \\ 2 + 7p_y + p_x(1 + 3p_y + 3p_x) \end{array}\right) \begin{array}{c} (1,1) \\ (1,2) \\ (1,3) \\ (2,1) \\ (2,2) \\ (2,3) \\ (3,1) \\ (3,2) \\ (3,3) \end{array}.
\end{equation}
Combining Eqs.~(\ref{eq_sensing1}) and (\ref{eq_sensing2}), we arrive at the joint transition probabilities $P(x,y\rightarrow x',y') = P(x',y'\, | \, x,y) \pi(x,y)$, which are used to perform the calculations in Sec. \ref{sec_sensing}.

\section{Convexity of local irreversibility}
\label{app_convex}

In order to compute $\dot{I}^{(k)}$, one must minimize the local irreversibility $\dot{I}$ subject to constraints on the $k^{\text{th}}$-order dynamics of the system. Here, we show that the local irreversibility is convex with respect to the transition probabilities $P(x\rightarrow x')$, and thus can be minimized using efficient techniques.

The gradient of the local irreversibility [Eq.~(\ref{eq_I})] is given by
\begin{equation}
\label{eq_grad}
\frac{\partial \dot{I}}{\partial P(x\rightarrow x')} = \log \frac{P(x\rightarrow x')}{P(x'\rightarrow x)} - \frac{P(x'\rightarrow x)}{P(x\rightarrow x')} + 1,
\end{equation}
where for simplicity $\log(\cdot)$ is natural logarithm. Since Eq.~(\ref{eq_grad}) only depends on $P(x\rightarrow x')$ and $P(x'\rightarrow x)$, we see that the Hessian of $\dot{I}$ takes the block diagonal form
\begin{equation}
H = \left(\begin{array}{ccc} \ddots & 0 & 0 \\ & \vspace{-9pt} & \\ 0 & H(x,x') & 0 \\ & \vspace{-15pt} & \\ 0 & 0 & \ddots \end{array} \right),
\end{equation}
where
\begin{align}
\hspace{-15pt} H(x,x') &= \left( \begin{array}{cc} \frac{\partial^2 \dot{I}}{\partial P(x\rightarrow x')^2} & \frac{\partial^2 \dot{I}}{\partial P(x\rightarrow x')\partial P(x'\rightarrow x)} \\ \vspace{-9pt} & \\ \frac{\partial^2 \dot{I}}{\partial P(x\rightarrow x') \partial P(x'\rightarrow x)} & \frac{\partial^2 \dot{I}}{\partial P(x'\rightarrow x)^2} \end{array} \right) \\
&= \left(\begin{array}{cc} \frac{P(x\rightarrow x') + P(x'\rightarrow x)}{P(x\rightarrow x')^2} & -\frac{P(x\rightarrow x') + P(x'\rightarrow x)}{P(x\rightarrow x')P(x'\rightarrow x)} \\ \vspace{-9pt} & \\ -\frac{P(x\rightarrow x') + P(x'\rightarrow x)}{P(x\rightarrow x')P(x'\rightarrow x)} & \frac{P(x\rightarrow x') + P(x'\rightarrow x)}{P(x'\rightarrow x)^2} \end{array} \right)
\end{align}
is the 2$\times$2 Hessian for the pair of states $(x,x')$. The eigenvalues of $H(x,x')$ are $\lambda_1 = \frac{(P(x\rightarrow x') + P(x'\rightarrow x))(P(x\rightarrow x')^2 + P(x'\rightarrow x)^2)}{P(x\rightarrow x')^2P(x'\rightarrow x)^2}$ and $\lambda_2 = 0$. Since $\lambda_1,\lambda_2 \ge 0$, and since the eigenvalues of $H$ are simply the eigenvalues of the different blocks $H(x,x')$ combined, we have established that $H$ is positive semidefinite, and therefore that the local irreversibility $\dot{I}$ is convex.

\section{Equivalence between independent and first--order irreversibilities}
\label{app_Sind}

Here we establish that the independent irreversibility $\dot{I}^{\text{ind}}$ is equivalent to the first--order minimum irreversibility $\dot{I}^{(1)}$. To do so, consider a hypothetical system $Q(x_{\rm i}\rightarrow x_{\rm i}',\, x_{-{\rm i}})$ that is consistent with the observed first--order dynamics $P(x_{\rm i}\rightarrow x_{\rm i}') = \sum_{x_{-{\rm i}}} P(x_{\rm i}\rightarrow x_{\rm i}',\, x_{-{\rm i}})$. Since $\dot{I}^{\text{ind}}(Q) = \dot{I}^{\text{ind}}(P)$, we have
\begin{align}
\label{eq_Iind2_1}
\dot{I}(Q) &= \dot{I}^{\text{ind}}(Q) + \dot{I}^{\text{int}}(Q) \\
& = \dot{I}^{\text{ind}}(P) + \dot{I}^{\text{int}}(Q) \\
\label{eq_Iind2_3}
&\ge \dot{I}^{\text{ind}}(P),
\end{align}
where the inequality follows from that fact that $\dot{I}^{\text{int}}(Q) \ge 0$. Thus, the independent irreversibility $\dot{I}^{\text{ind}}(P)$ is a lower bound on the local irreversibility $\dot{I}(Q)$ of any hypothetical system $Q$ consistent with the observed first--order dynamics. Since the first--order irreversibility $\dot{I}^{(1)}$ is just the minimum of $\dot{I}(Q)$ among all such systems $Q$, we have found that $\dot{I}^{(1)} \ge \dot{I}^{\text{ind}}$.

In order to establish that $\dot{I}^{(1)} = \dot{I}^{\text{ind}}$, all that remains is to identify a hypothetical system $Q$ that achieves the lower bound in Eqs.~(\ref{eq_Iind2_1}--\ref{eq_Iind2_3}). Specifically, we seek a system $Q$ that is consistent with the observed first--order dynamics, yet has interaction irreversibility $\dot{I}^{\text{int}}(Q) = 0$. Consider, for example, a system $Q$ in which the dynamics of each element $\rm i$ are independent from the rest of the system, such that $Q(x_{\rm i}\rightarrow x_{\rm i}',\, x_{-{\rm i}}) = Q(x_{\rm i} \rightarrow x_{\rm i}') = P(x_{\rm i} \rightarrow x_{\rm i}')$ for all $x_{-{\rm i}}$. Using Eqs.~(\ref{eq_Iint_1}--\ref{eq_Iint_3}), one can verify that such a system has zero interaction irreversibility, thereby saturating the lower bound in Eqs.~(\ref{eq_Iind2_1}--\ref{eq_Iind2_3}). We have therefore shown that $\dot{I}^{(1)}$ (the minimum local irreversibility consistent with first--order dynamics) is equivalent to the independent irreversibility $\dot{I}^{\text{ind}}$.

\section{Noisy logical functions}
\label{app_logical}

In Sec. \ref{sec_logical}, we examine a system of three binary variables: two inputs $x$ and $y$ that flip with probability $p_{\text{flip}}$, and an output variable $z$ that performs a logical function on $x$ and $y$, but with error rate $p_{\text{error}}$ [see Fig.~\ref{fig_logical_1}(a)]. Specifically, the dynamics of the input variables are defined by the conditional transition probabilities
\begin{equation}
P(x'\, |\, x) = P(y'\, |\, y) = \left(\begin{array}{cc} 1- p_{\text{flip}} & p_{\text{flip}} \\ p_{\text{flip}} & 1 - p_{\text{flip}} \end{array} \right),
\end{equation}
and the dynamics of the output variable are defined by
\begin{equation}
P(z' \, | \, x,y) = \left\{\begin{array}{cc} 1-p_{\text{error}}, & z' = f(x,y) \\ p_{\text{error}}, & z' \neq f(x,y) \end{array} \right. ,
\end{equation}
where $f(x,y)$ is the logical function performed by $z$. Randomly picking one variable to update at each point in time, the conditional transition probabilities for the entire system are given by
\begin{align}
\label{eq_Plogical}
&P(x',y',z' \, | \, x, y,z) = \frac{1}{3}\big( P(x'\, |\, x)\delta_{y,y'}\delta_{z,z'} \\
&\quad\quad\quad\quad\quad\quad + P(y'\, |\, y)\delta_{x,x'}\delta_{z,z'} + P(z'\, |\, x,y)\delta_{x,x'}\delta_{y,y'} \big). \nonumber
\end{align}
Using Eq.~(\ref{eq_Plogical}), one can solve for the stationary distribution $\pi(x,y,z)$ and then compute the joint transition probabilities $P(x,y,z\rightarrow x',y',z') = P(x',y',z' \, | \, x, y,z)\pi(x,y,z)$.

\section{Independent irreversibility vanishes for binary steady--state systems}
\label{app_bin}

For binary steady--state systems, such as the logical functions in Sec. \ref{sec_logical} and the neurons in Sec. \ref{sec_neuron}, the independent irreversibility $\dot{I}^{\text{ind}}$ is zero. To see this, note that the marginal dynamics of any binary steady--state variable are defined by the conditional transition probabilities
\begin{equation}
P(x_{\rm i}'\, | \, x_{\rm i}) = \left(\begin{array}{cc} 1-p_{\rm i} & p_{\rm i} \\ q_{\rm i} & 1-q_{\rm i} \end{array} \right),
\end{equation}
where $0 \le p_{\rm i}, q_{\rm i} \le 1$ are the probabilities of $\rm i$ switching between its two states. The marginal steady--state distribution for $\rm i$ is $\pi(x_{\rm i}) = \frac{1}{p_{\rm i}+q_{\rm i}} (q_{\rm i}, p_{\rm i})^T$, and thus the marginal joint transition probabilities are given by
\begin{align}
P(x_{\rm i}\rightarrow x_{\rm i}') &= P(x_{\rm i}'\, |\, x_{\rm i})\pi(x_{\rm i}) \\
&= \frac{1}{p_{\rm i} + q_{\rm i}} \left(\begin{array}{cc} (1-p_{\rm i})q_{\rm i} & p_{\rm i}q_{\rm i} \\ q_{\rm i}p_{\rm i} & (1-q_{\rm i})p_{\rm i} \end{array} \right).
\end{align}
Since the above transition probabilities are symmetric, the marginal dynamics of each element $\rm i$ obey detailed balance (such that $\dot{I}^{\text{ind}}_{\rm i} = 0$). Thus, we find that the independent irreversibility of the entire system $\dot{I}^{\text{ind}} = \sum_{{\rm i}=1}^N \dot{I}^{\text{ind}}_{\rm i}$ is zero. We emphasize that this only holds for the \textit{local} irreversibility; if we consider non--Markov effects in strings of 3, 4, or more points in time, then binary steady--state variables can break time-reversal symmetry and define an arrow of time \cite{Roldan-01}.

\section{Neuronal recordings}
\label{app_neuron}

The neuronal data examined in Sec. \ref{sec_neuron} were recorded from larval tiger salamander retina, which were dissected, perfused with Ringer's solution, and pressed onto dense arrays of 252 electrodes with 30--$\mu$m spacing, as described in Ref. \cite{Palmer-01}. In the experiments, which lasted 4--6 hours, movies were projected onto the photoreceptor layer of the retina via an objective lens, and voltages were recorded at 10 kHz. Spikes were sorted conservatively (as described in Ref. \cite{Marre-01}), yielding 53 reliable cells from which groups were randomly selected for analysis.

The stimuli were presented on a 360$\times$600 display, with pixels of size $3.81 \,\mu {\rm m}$ on the retina and a frame rate of 60 frames per second. All stimuli were normalized to the same average light intensity. The natural movie depicted a fish swimming in a tank, repeated 102 times. The moving bar was 11 pixels wide and black on a gray background, with trajectories displayed 62 times. The trajectory of the bar's vertical position was generated by a stochastic process equivalent to a Brownian particle on a spring attached to the center of the display. Specifically, the vertical position $x_t$ and velocity $v_t$ of the bar were updated at each time $t$ according to the equations of motion:
\begin{equation}
x_{t+\tau} = x_t + v_t \tau,
\end{equation}
and
\begin{equation}
v_{t + \tau} = (1 - \Gamma \tau)v_t - \omega^2 x_t \tau + \xi_t \sqrt{D \tau},
\end{equation}
where $\tau = 1/60\,{\rm s}$ is the time step (which matches the frame rate of the visual display), $\omega = 3\pi \, {\rm s}^{-1}$ is the natural frequency, $\Gamma = 20\, {\rm s}^{-1}$ parameterizes the damping (chosen such that the dynamics are slightly overdamped), and $D = 2.7\times 10^6 \, {\rm pixel}^2 / {\rm s}^3$ is chosen to allow reasonable range of motion. For the repeated bar stimulus, the same trajectory was repeated 62 times.

\section{Correcting for finite data}
\label{app_finite}

For time--series data, such as the neuronal spiking examined in Sec. \ref{sec_neuron}, in order to estimate quantities of interest---such as transition rates, flux rates, and changes in state probabilities (Fig.~\ref{fig_neuron_1}); local irreversibilities $\dot{I}$ (Fig.~\ref{fig_neuron_2}); and interaction irreversibilities $\dot{I}^{(k)}_{\text{int}}$ (Fig.~\ref{fig_neuron_3})---one must correct for finite--data effects \cite{Strong-01, Schneidman-01, Palmer-01}. To do so, for a given stimulus and group of neurons, we begin with a list of the observed transitions $\{x(t) \rightarrow x(t+1)\}$. For a given set of data fractions $f$, we subsample the transitions (without replacement) in a hierarchical fashion, such that each subsample of transitions is a subset of the larger subsamples. For the neuronal data in Sec. \ref{sec_neuron}, we find that data fractions $f = \{1, 0.9, 0.8, 0.7, 0.6, 0.5\}$ are sufficient.

\begin{figure}
\centering
\includegraphics[width = \columnwidth]{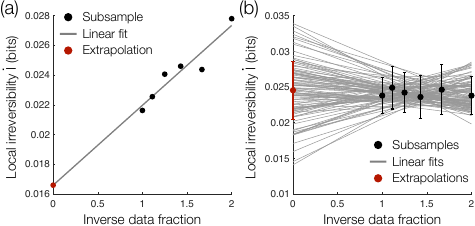} \\
\caption{Correcting for finite--data effects on local irreversibility. (a) Estimated local irreversibility $\dot{I}$ (black markers) versus inverse data fraction for one group of $N=5$ neurons responding to a natural movie. Grey line indicates a linear fit, and red marker indicates the extrapolation to infinite data. (b) Estimated local irreversibility (black markers), linear fits (grey lines), and extrapolation to infinite data (red marker) after repeating the process in panel (a) 100 times for the same group of neurons. Data points and error bars reflect averages and standard deviations over the 100 repetitions. \label{fig_finite}}
\end{figure}

For each data fraction $f$, we estimate the quantity of interest. For example, for the local irreversibility, we use the estimate
\begin{equation}
\label{eq_Iest}
\dot{I} = \sum_{x,x'} \tilde{P}(x\rightarrow x') \log \frac{\tilde{P}(x\rightarrow x')}{\tilde{P}(x'\rightarrow x)},
\end{equation}
where
\begin{equation}
\tilde{P}(x\rightarrow x') = \frac{N(x\rightarrow x') + 1}{\sum_{y,y'} \left(N(y\rightarrow y') + 1\right)}
\end{equation}
are the maximum likelihood probabilities with one pseudocount for each transition, and $N(x\rightarrow x')$ is the number of times that the transition $x\rightarrow x'$ was observed in the data. We include pseudocounts to avoid infinities in Eq.~\ref{eq_Iest}, but we confirm that the na\"{i}ve estimator without pseudocounts yields the same results. After estimating the quantity of interest for all fractions $f$, we then extrapolate to the infinite--data limit using a linear fit with respect to the inverse data fraction $1/f$ [Fig.~\ref{fig_finite}(a)]. Repeating this process 100 times, we arrive at both an average and standard deviation for the infinite--data estimates of the desired quantity [Fig.~\ref{fig_finite}(b)].

\begin{figure}
\centering
\includegraphics[width = \columnwidth]{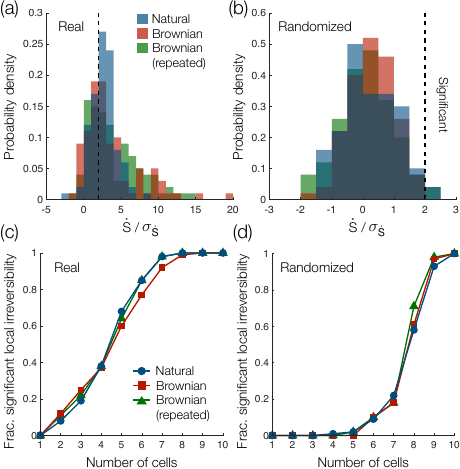} \\
\caption{Estimated local irreversibilities of real and time--shuffled data. (a--b) Distributions of local irreversibility estimates $\dot{I}$, normalized by standard deviations $\sigma_{\dot{I}}$, for real time series (a) and after randomizing spike times (b). Distributions are over the same 100 groups of $N=5$ cells analyzed in Figs.~\ref{fig_neuron_2} and \ref{fig_neuron_3}, and the dashed line indicates the threshold for significance. (c--d) Fraction of cell groups with significant local irreversibility as a function of group size $N$ for real time series (c) and after randomizing spike times (d). For each size, fractions are computed over 100 random groups. \label{fig_shuffle}}
\end{figure}

To check that the above procedure gives accurate estimates for the local irreversibility $\dot{I}$, we note that randomizing the timing of spikes should destroy the arrow of time. Thus, for time--randomized data, the estimated local irreversibility should vanish in the infinite--data limit. Consider the 100 groups of $N=5$ neurons analyzed in Figs.~\ref{fig_neuron_2} and \ref{fig_neuron_3}. Among these groups, after correcting for finite--data effects, we find that $60$--$68\%$ exhibit significant local irreversibility $\dot{I}$, depending on the stimulus [Fig.~\ref{fig_shuffle}(a)]. By contrast, after randomizing the spike times, the local irreversibility estimates are centered around zero, with only $0$--$2\%$ of groups exhibiting significant local irreversibility [Fig.~\ref{fig_shuffle}(b)]. Examining different group sizes, we find that the percentage of groups with significant local irreversibility increases from $\sim$10\% for $N=2$ cells to $\sim$100\% for $N \ge 8$ cells [Fig.~\ref{fig_shuffle}(c)]. Importantly, after randomizing spike times, groups of $N \le 5$ cells are almost always locally reversible, as desired [Fig.~\ref{fig_shuffle}(d)]. However, even for time--randomized data, we find that some groups of $N \ge 6$ cells exhibit significant local irreversibility [Fig.~\ref{fig_shuffle}(d)], demonstrating finite--data effects cannot be adequately accounted for. We therefore conclude that $N = 5$ is the largest number of cells for which we can consistently estimate local irreversibility $\dot{I}$ in our dataset.

\section{Minimizing local irreversibility}
\label{app_min}

Computing the $k^{\text{th}}$--order minimum irreversibility $\dot{I}^{(k)}$ requires finding a hypothetical system $Q(x_{\rm i}\rightarrow x_{\rm i}',\, x_{-{\rm i}})$ that matches the observed $k^{\text{th}}$--order marginal dynamics, but otherwise has minimum local irreversibility $\dot{I}(Q)$. We remark that the local irreversibility $\dot{I}$ is convex (see Appendix \ref{app_convex}), and thus computing $\dot{I}^{(k)}$ is a constrained convex minimization problem for which there exist efficient optimization methods. From a practical perspective, there are only two main hurdles to overcome: (i) adapting an existing convex minimization technique for our problem, and (ii) writing down the constraints on the $k^{\text{th}}$--order dynamics. Here, we address these challenges for binary systems.

\subsection{Frank--Wolfe algorithm}

To minimize the local irreversibility $\dot{I}$ given a set of constraints, we employ the Frank--Wolfe algorithm, which efficiently converges to a local (and therefore global) minimum. Specifically, we initialize $Q$ using any dynamics that match the observed constraints (for example, one can begin with the observed dynamics $Q = P$). We then iterate the following steps:
\begin{enumerate}
\item First, we compute the gradient of the local irreversibility:
\begin{align}
\frac{\partial \dot{I}}{\partial Q(x_{\rm i}\rightarrow x_{\rm i}',\, x_{-{\rm i}})} &= \log \frac{Q(x_{\rm i}\rightarrow x_{\rm i}',\, x_{-{\rm i}})}{Q(x_{\rm i}' \rightarrow x_{\rm i},\, x_{-{\rm i}})} \\
& \quad\quad\quad - \frac{Q(x_{\rm i}' \rightarrow x_{\rm i},\, x_{-{\rm i}})}{Q(x_{\rm i}\rightarrow x_{\rm i}',\, x_{-{\rm i}})} + 1, \nonumber
\end{align}
where $\log(\cdot)$ represents the natural logarithm for simplicity.
\item Second, we solve for the dynamics $Q^*$ that obey the desired constraints while minimizing the inner product with the gradient:
\begin{equation}
\sum_{{\rm i}=1}^N \sum_{x_{-{\rm i}}} \sum_{x_{\rm i}, x_{\rm i}'} Q(x_{\rm i}\rightarrow x_{\rm i}',\, x_{-{\rm i}}) \frac{\partial \dot{I}}{\partial Q(x_{\rm i}\rightarrow x_{\rm i}',\, x_{-{\rm i}})}.
\end{equation}
We note that this constrained linear minimization problem is a linear program, and thus can be efficiently solved using standard techniques (e.g., the \texttt{linprog} function in MATLAB).
\item Finally, we take a step toward $Q^*$, such that $Q \leftarrow Q + \alpha_t(Q^* - Q)$, where $\alpha_t = \alpha_0/t$ is the step size, which decreases with the number of iterations $t$.
\end{enumerate}
An implementation of the above algorithm is available at \texttt{github.com/ChrisWLynn/Decompose{\_}irreversibility}.

\subsection{Constraining $k^{\text{th}}$--order dynamics}

We seek to constrain the $k^{\text{th}}$--order dynamics of a binary, multipartite system with joint transition probabilities $P(x_{\rm i}\rightarrow x_{\rm i}',\, x_{-{\rm i}})$. For each element $\rm i$, consider a group of $k-1$ of the remaining elements $K \subseteq \{1, \hdots, {\rm i}-1, {\rm i}+1, \hdots, N\}$. Let $x_K$ denote the states of the elements in $K$, and $x_{-\{{\rm i},K\}}$ the states of the elements not in $\rm i$ nor $K$. The marginal dynamics of $\rm i$ with the elements in $K$ held fixed are then given by
\begin{equation}
P(x_{\rm i}\rightarrow x_{\rm i}',\, x_K) = \sum_{x_{-\{{\rm i}, K\}}} P(x_{\rm i} \rightarrow x_{\rm i}',\, x_{-{\rm i}}).
\end{equation}
Constraining the $k^{\text{th}}$--order dynamics amounts to constraining the marginal probabilities $P(x_{\rm i}\rightarrow x_{\rm i}',\, x_K)$ for all elements $\rm i$ and all groups of the remaining elements $K$ of size $k-1$. For example, if $K$ is empty, then we arrive at the independent (first--order) dynamics $P(x_{\rm i}\rightarrow x_{\rm i}')$. If $K$ consists of one element $\rm j$, then we have the pairwise (second--order) dynamics $P(x_{\rm i}\rightarrow x_{\rm i}',\, x_{\rm j})$ discussed in Sec. \ref{sec_dec}. We remark, however, that these marginal probabilities are not all independent, and therefore the set of constraints is overdetermined.

To write down independent constraints that fully define the $k^{\text{th}}$--order dynamics, it helps to consider an analogy with Ising systems. Consider a binary system with state probabilities $P(x)$. It is known that the $k^{\text{th}}$--order marginal probabilities $P(x_K) = \sum_{x_{-K}} P(x)$ are completely defined by the correlations between groups of elements up to size $k$: $\left<x_{\rm i}\right>$, $\left<x_{\rm i}x_{\rm j}\right>$, $\hdots$, $\left<\prod_{{\rm i}\in K}x_{\rm i}\right>$, where $\left<\cdot\right>$ represents an average over $P(x)$ \cite{Schneidman-02}. Moreover, these correlations are independent, thus forming a basis for the $k^{\text{th}}$--order probabilities $P(x_K)$.

Here, we wish to constrain the $k^{\text{th}}$--order transition probabilities $P(x_{\rm i}\rightarrow x_{\rm i}',\, x_K)$ for all elements $\rm i$ and all groups of the remaining elements $K$ of size $k-1$. For a given transition $x_{\rm i}\rightarrow x_{\rm i}'$, we denote the correlation between a set of the remaining elements $K$ by
\begin{equation}
\Big<\prod_{{\rm j} \in K} x_{\rm j}\Big>_{x_{\rm i}\rightarrow x_{\rm i}'} = \sum_{x_{-{\rm i}}} \prod_{{\rm j} \in K} x_{\rm j} P(x_{\rm i}\rightarrow x_{\rm i}', \, x_{-{\rm i}}).
\end{equation}
If $K$ is empty, then we simply arrive at the independent transition probabilities:
\begin{equation}
\left< 1 \right>_{x_{\rm i}\rightarrow x_{\rm i}'} = \sum_{x_{-{\rm i}}} P(x_{\rm i}\rightarrow x_{\rm i}', \, x_{-{\rm i}}) = P(x_{\rm i}\rightarrow x_{\rm i}').
\end{equation}
By analogy with Ising systems, for each transition $x_{\rm i}\rightarrow x_{\rm i}'$, the $k^{\text{th}}$--order marginal probabilities $P(x_{\rm i}\rightarrow x_{\rm i}',\, x_K)$ can be defined by the correlations between groups of elements (not including $\rm i$) from the empty set up to size $k-1$: $\left<1\right>_{x_{\rm i}\rightarrow x_{\rm i}'}$, $\left<x_{\rm j}\right>_{x_{\rm i}\rightarrow x_{\rm i}'}$, $\hdots$, $\left<\prod_{{\rm j}\in K}x_{\rm j}\right>_{x_{\rm i}\rightarrow x_{\rm i}'}$. We can then constrain the $k^{\text{th}}$--order dynamics of the entire system by computing the above correlations for each of the $2N$ transitions $x_{\rm i}\rightarrow x_{\rm i}'$ (not including self--transitions). We remark that we do not need to constrain self--transitions $x\rightarrow x$ because they do not contribute to the local irreversibility [Eq.~\ref{eq_I}]. Code for constraining the $k^{\text{th}}$--order dynamics of binary, multipartite systems is available at \texttt{github.com/ChrisWLynn/Decompose{\_}irreversibility}.

\section{Groups of neurons operate at steady state}
\label{app_ss}

\begin{figure}
\centering
\includegraphics[width = \columnwidth]{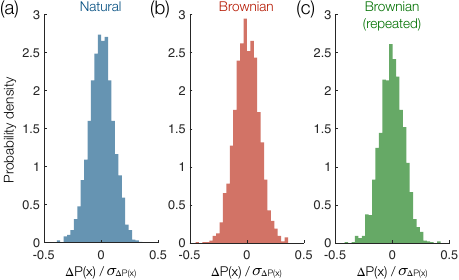} \\
\caption{Neurons operate at stochastic steady states. (a--c) Distributions of changes in state probabilities $\Delta P(x) = \sum_{x'} P(x'\rightarrow x) - P(x \rightarrow x')$, normalized by the standard deviation $\sigma_{\Delta P(x)}$, for groups of $N=5$ cells responding to a natural movie (a), moving bar stimulus (b), and a repeated moving bar stimulus (c). Distributions are over the $2^N=32$ different states for the 100 random groups analyzed in Figs.~\ref{fig_neuron_2} and \ref{fig_neuron_3}. \label{fig_ss}}
\end{figure}

In Figure \ref{fig_neuron_1}, we see that a group of $N=3$ neurons operates at a non--equilibrium steady state. Here, we demonstrate that steady--state dynamics are not specific just to this group, but are instead a general feature of all groups of neurons analyzed in this paper. To determine if a system operates at steady state, one must examine whether its state probabilities are stationary in time. The change in the probability $P(x)$ of a state $x$ during one time step is given by $\Delta P(x) = \sum_{x'} P(x'\rightarrow x) - P(x \rightarrow x')$. In steady state this should be zero, but more precisely we expect that it will be a random number with a variance set by the errors in sampling the underlying distributions. In Fig.~\ref{fig_ss}, we plot the distributions of $\Delta P(x)$, normalized by the relevant standard deviation $\sigma_{\Delta P(x)}$, for the groups of $N=5$ cells analyzed in Figs.~\ref{fig_neuron_2} and \ref{fig_neuron_3}. We note that these quantities are estimated using the same finite--data correction techniques described in Appendix \ref{app_finite}. Across all stimuli, we find that the changes in state probabilities $\Delta P(x)$ for all cell groups are small relative to errors; that is, for all stimuli, all groups of neurons appear to operate at steady state.

\section{Data and code availability}
\label{app_data}

The data and code used to perform the analyses in this paper are openly available at \\ \texttt{github.com/ChrisWLynn/Decompose{\_}irreversibility}.


\bibliography{DetailedBalanceBib}

\begin{thebibliography}{43}%
\makeatletter
\providecommand \@ifxundefined [1]{%
 \@ifx{#1\undefined}
}%
\providecommand \@ifnum [1]{%
 \ifnum #1\expandafter \@firstoftwo
 \else \expandafter \@secondoftwo
 \fi
}%
\providecommand \@ifx [1]{%
 \ifx #1\expandafter \@firstoftwo
 \else \expandafter \@secondoftwo
 \fi
}%
\providecommand \natexlab [1]{#1}%
\providecommand \enquote  [1]{``#1''}%
\providecommand \bibnamefont  [1]{#1}%
\providecommand \bibfnamefont [1]{#1}%
\providecommand \citenamefont [1]{#1}%
\providecommand \href@noop [0]{\@secondoftwo}%
\providecommand \href [0]{\begingroup \@sanitize@url \@href}%
\providecommand \@href[1]{\@@startlink{#1}\@@href}%
\providecommand \@@href[1]{\endgroup#1\@@endlink}%
\providecommand \@sanitize@url [0]{\catcode `\\12\catcode `\$12\catcode
  `\&12\catcode `\#12\catcode `\^12\catcode `\_12\catcode `\%12\relax}%
\providecommand \@@startlink[1]{}%
\providecommand \@@endlink[0]{}%
\providecommand \url  [0]{\begingroup\@sanitize@url \@url }%
\providecommand \@url [1]{\endgroup\@href {#1}{\urlprefix }}%
\providecommand \urlprefix  [0]{URL }%
\providecommand \Eprint [0]{\href }%
\providecommand \doibase [0]{http://dx.doi.org/}%
\providecommand \selectlanguage [0]{\@gobble}%
\providecommand \bibinfo  [0]{\@secondoftwo}%
\providecommand \bibfield  [0]{\@secondoftwo}%
\providecommand \translation [1]{[#1]}%
\providecommand \BibitemOpen [0]{}%
\providecommand \bibitemStop [0]{}%
\providecommand \bibitemNoStop [0]{.\EOS\space}%
\providecommand \EOS [0]{\spacefactor3000\relax}%
\providecommand \BibitemShut  [1]{\csname bibitem#1\endcsname}%
\let\auto@bib@innerbib\@empty
\bibitem [{\citenamefont {Parrondo}\ \emph {et~al.}(2015)\citenamefont
  {Parrondo}, \citenamefont {Horowitz},\ and\ \citenamefont
  {Sagawa}}]{Parrondo-01}%
  \BibitemOpen
  \bibfield  {author} {\bibinfo {author} {\bibfnamefont {Juan~MR}\ \bibnamefont
  {Parrondo}}, \bibinfo {author} {\bibfnamefont {Jordan~M}\ \bibnamefont
  {Horowitz}}, \ and\ \bibinfo {author} {\bibfnamefont {Takahiro}\ \bibnamefont
  {Sagawa}},\ }\bibfield  {title} {\enquote {\bibinfo {title} {Thermodynamics
  of information},}\ }\href@noop {} {\bibfield  {journal} {\bibinfo  {journal}
  {Nat. Phys.}\ }\textbf {\bibinfo {volume} {11}},\ \bibinfo {pages} {131--139}
  (\bibinfo {year} {2015})}\BibitemShut {NoStop}%
\bibitem [{\citenamefont {Peliti}\ and\ \citenamefont
  {Pigolotti}(2021)}]{Peliti-01}%
  \BibitemOpen
  \bibfield  {author} {\bibinfo {author} {\bibfnamefont {Luca}\ \bibnamefont
  {Peliti}}\ and\ \bibinfo {author} {\bibfnamefont {Simone}\ \bibnamefont
  {Pigolotti}},\ }\href@noop {} {\emph {\bibinfo {title} {Stochastic
  Thermodynamics: An Introduction}}}\ (\bibinfo  {publisher} {Princeton
  University Press},\ \bibinfo {year} {2021})\BibitemShut {NoStop}%
\bibitem [{\citenamefont {Gnesotto}\ \emph {et~al.}(2018)\citenamefont
  {Gnesotto}, \citenamefont {Mura}, \citenamefont {Gladrow},\ and\
  \citenamefont {Broedersz}}]{Gnesotto-01}%
  \BibitemOpen
  \bibfield  {author} {\bibinfo {author} {\bibfnamefont {F~S}\ \bibnamefont
  {Gnesotto}}, \bibinfo {author} {\bibfnamefont {Federica}\ \bibnamefont
  {Mura}}, \bibinfo {author} {\bibfnamefont {Jannes}\ \bibnamefont {Gladrow}},
  \ and\ \bibinfo {author} {\bibfnamefont {C~P}\ \bibnamefont {Broedersz}},\
  }\bibfield  {title} {\enquote {\bibinfo {title} {Broken detailed balance and
  non-equilibrium dynamics in living systems: {A} review},}\ }\href@noop {}
  {\bibfield  {journal} {\bibinfo  {journal} {Rep. Prog. Phys.}\ }\textbf
  {\bibinfo {volume} {81}},\ \bibinfo {pages} {066601} (\bibinfo {year}
  {2018})}\BibitemShut {NoStop}%
\bibitem [{\citenamefont {Lynn}\ \emph {et~al.}()\citenamefont {Lynn},
  \citenamefont {Holmes}, \citenamefont {Bialek},\ and\ \citenamefont
  {Schwab}}]{Lynn-11}%
  \BibitemOpen
  \bibfield  {author} {\bibinfo {author} {\bibfnamefont {Christopher~W}\
  \bibnamefont {Lynn}}, \bibinfo {author} {\bibfnamefont {Caroline~M}\
  \bibnamefont {Holmes}}, \bibinfo {author} {\bibfnamefont {William}\
  \bibnamefont {Bialek}}, \ and\ \bibinfo {author} {\bibfnamefont {David~J}\
  \bibnamefont {Schwab}},\ }\bibfield  {title} {\enquote {\bibinfo {title}
  {Decomposing the local arrow of time in interacting systems},}\ }\href@noop
  {} {\bibinfo  {journal} {Preprint: arxiv.org/abs/2112.14721}\ }\BibitemShut
  {NoStop}%
\bibitem [{\citenamefont {Schneidman}\ \emph {et~al.}(2003)\citenamefont
  {Schneidman}, \citenamefont {Still}, \citenamefont {Berry~II},\ and\
  \citenamefont {Bialek}}]{Schneidman-02}%
  \BibitemOpen
\bibfield  {journal} {  }\bibfield  {author} {\bibinfo {author} {\bibfnamefont
  {E}~\bibnamefont {Schneidman}}, \bibinfo {author} {\bibfnamefont
  {S}~\bibnamefont {Still}}, \bibinfo {author} {\bibfnamefont {M.~J.}\
  \bibnamefont {Berry~II}}, \ and\ \bibinfo {author} {\bibfnamefont
  {W}~\bibnamefont {Bialek}},\ }\bibfield  {title} {\enquote {\bibinfo {title}
  {Network information and connected correlations},}\ }\href@noop {} {\bibfield
   {journal} {\bibinfo  {journal} {Phys. Rev. Lett.}\ }\textbf {\bibinfo
  {volume} {91}},\ \bibinfo {pages} {238701} (\bibinfo {year}
  {2003})}\BibitemShut {NoStop}%
\bibitem [{\citenamefont {Marre}\ \emph {et~al.}(2012)\citenamefont {Marre},
  \citenamefont {Amodei}, \citenamefont {Deshmukh}, \citenamefont {Sadeghi},
  \citenamefont {Soo}, \citenamefont {Holy},\ and\ \citenamefont
  {Berry~II}}]{Marre-01}%
  \BibitemOpen
  \bibfield  {author} {\bibinfo {author} {\bibfnamefont {O}~\bibnamefont
  {Marre}}, \bibinfo {author} {\bibfnamefont {D}~\bibnamefont {Amodei}},
  \bibinfo {author} {\bibfnamefont {N}~\bibnamefont {Deshmukh}}, \bibinfo
  {author} {\bibfnamefont {K}~\bibnamefont {Sadeghi}}, \bibinfo {author}
  {\bibfnamefont {F}~\bibnamefont {Soo}}, \bibinfo {author} {\bibfnamefont
  {TE}~\bibnamefont {Holy}}, \ and\ \bibinfo {author} {\bibfnamefont {M~J}\
  \bibnamefont {Berry~II}},\ }\bibfield  {title} {\enquote {\bibinfo {title}
  {Mapping a complete neural population in the retina},}\ }\href@noop {}
  {\bibfield  {journal} {\bibinfo  {journal} {J. Neurosci.}\ }\textbf {\bibinfo
  {volume} {32}},\ \bibinfo {pages} {14859--14873} (\bibinfo {year}
  {2012})}\BibitemShut {NoStop}%
\bibitem [{\citenamefont {Palmer}\ \emph {et~al.}(2015)\citenamefont {Palmer},
  \citenamefont {Marre}, \citenamefont {Berry~II},\ and\ \citenamefont
  {Bialek}}]{Palmer-01}%
  \BibitemOpen
  \bibfield  {author} {\bibinfo {author} {\bibfnamefont {S~E}\ \bibnamefont
  {Palmer}}, \bibinfo {author} {\bibfnamefont {O}~\bibnamefont {Marre}},
  \bibinfo {author} {\bibfnamefont {M~J}\ \bibnamefont {Berry~II}}, \ and\
  \bibinfo {author} {\bibfnamefont {W}~\bibnamefont {Bialek}},\ }\bibfield
  {title} {\enquote {\bibinfo {title} {Predictive information in a sensory
  population},}\ }\href@noop {} {\bibfield  {journal} {\bibinfo  {journal}
  {Proc. Natl. Acad. Sci.}\ }\textbf {\bibinfo {volume} {112}},\ \bibinfo
  {pages} {6908--6913} (\bibinfo {year} {2015})}\BibitemShut {NoStop}%
\bibitem [{\citenamefont {Cover}\ and\ \citenamefont
  {Thomas}(2012)}]{Cover-01}%
  \BibitemOpen
  \bibfield  {author} {\bibinfo {author} {\bibfnamefont {T~M}\ \bibnamefont
  {Cover}}\ and\ \bibinfo {author} {\bibfnamefont {J~A}\ \bibnamefont
  {Thomas}},\ }\href@noop {} {\emph {\bibinfo {title} {Elements of Information
  Theory}}}\ (\bibinfo  {publisher} {John Wiley \& Sons},\ \bibinfo {year}
  {2012})\BibitemShut {NoStop}%
\bibitem [{\citenamefont {Seifert}(2005)}]{Seifert-01}%
  \BibitemOpen
  \bibfield  {author} {\bibinfo {author} {\bibfnamefont {Udo}\ \bibnamefont
  {Seifert}},\ }\bibfield  {title} {\enquote {\bibinfo {title} {Entropy
  production along a stochastic trajectory and an integral fluctuation
  theorem},}\ }\href@noop {} {\bibfield  {journal} {\bibinfo  {journal} {Phys.
  Rev. Lett.}\ }\textbf {\bibinfo {volume} {95}},\ \bibinfo {pages} {040602}
  (\bibinfo {year} {2005})}\BibitemShut {NoStop}%
\bibitem [{\citenamefont {Ge}\ and\ \citenamefont {Qian}(2013)}]{Ge-01}%
  \BibitemOpen
  \bibfield  {author} {\bibinfo {author} {\bibfnamefont {Hao}\ \bibnamefont
  {Ge}}\ and\ \bibinfo {author} {\bibfnamefont {Hong}\ \bibnamefont {Qian}},\
  }\bibfield  {title} {\enquote {\bibinfo {title} {Dissipation, generalized
  free energy, and a self-consistent nonequilibrium thermodynamics of
  chemically driven open subsystems},}\ }\href@noop {} {\bibfield  {journal}
  {\bibinfo  {journal} {Phys. Rev. E}\ }\textbf {\bibinfo {volume} {87}},\
  \bibinfo {pages} {062125} (\bibinfo {year} {2013})}\BibitemShut {NoStop}%
\bibitem [{\citenamefont {Horowitz}\ and\ \citenamefont
  {Esposito}(2014)}]{Horowitz-01}%
  \BibitemOpen
  \bibfield  {author} {\bibinfo {author} {\bibfnamefont {J.~M.}\ \bibnamefont
  {Horowitz}}\ and\ \bibinfo {author} {\bibfnamefont {M}~\bibnamefont
  {Esposito}},\ }\bibfield  {title} {\enquote {\bibinfo {title} {Thermodynamics
  with continuous information flow},}\ }\href@noop {} {\bibfield  {journal}
  {\bibinfo  {journal} {Phys. Rev. X}\ }\textbf {\bibinfo {volume} {4}},\
  \bibinfo {pages} {031015} (\bibinfo {year} {2014})}\BibitemShut {NoStop}%
\bibitem [{\citenamefont {Wolpert}(2020)}]{Wolpert-01}%
  \BibitemOpen
  \bibfield  {author} {\bibinfo {author} {\bibfnamefont {D~H}\ \bibnamefont
  {Wolpert}},\ }\bibfield  {title} {\enquote {\bibinfo {title} {Minimal entropy
  production rate of interacting systems},}\ }\href@noop {} {\bibfield
  {journal} {\bibinfo  {journal} {New J. Phys.}\ }\textbf {\bibinfo {volume}
  {22}},\ \bibinfo {pages} {113013} (\bibinfo {year} {2020})}\BibitemShut
  {NoStop}%
\bibitem [{\citenamefont {Lan}\ \emph {et~al.}(2012)\citenamefont {Lan},
  \citenamefont {Sartori}, \citenamefont {Neumann}, \citenamefont {Sourjik},\
  and\ \citenamefont {Tu}}]{Lan-01}%
  \BibitemOpen
  \bibfield  {author} {\bibinfo {author} {\bibfnamefont {Ganhui}\ \bibnamefont
  {Lan}}, \bibinfo {author} {\bibfnamefont {Pablo}\ \bibnamefont {Sartori}},
  \bibinfo {author} {\bibfnamefont {Silke}\ \bibnamefont {Neumann}}, \bibinfo
  {author} {\bibfnamefont {Victor}\ \bibnamefont {Sourjik}}, \ and\ \bibinfo
  {author} {\bibfnamefont {Yuhai}\ \bibnamefont {Tu}},\ }\bibfield  {title}
  {\enquote {\bibinfo {title} {The energy--speed--accuracy trade-off in sensory
  adaptation},}\ }\href@noop {} {\bibfield  {journal} {\bibinfo  {journal}
  {Nat. Phys.}\ }\textbf {\bibinfo {volume} {8}},\ \bibinfo {pages} {422}
  (\bibinfo {year} {2012})}\BibitemShut {NoStop}%
\bibitem [{\citenamefont {Barato}\ \emph {et~al.}(2013)\citenamefont {Barato},
  \citenamefont {Hartich},\ and\ \citenamefont {Seifert}}]{Barato-01}%
  \BibitemOpen
  \bibfield  {author} {\bibinfo {author} {\bibfnamefont {A.~C.}\ \bibnamefont
  {Barato}}, \bibinfo {author} {\bibfnamefont {D}~\bibnamefont {Hartich}}, \
  and\ \bibinfo {author} {\bibfnamefont {U}~\bibnamefont {Seifert}},\
  }\bibfield  {title} {\enquote {\bibinfo {title} {Information-theoretic versus
  thermodynamic entropy production in autonomous sensory networks},}\
  }\href@noop {} {\bibfield  {journal} {\bibinfo  {journal} {Phys. Rev. E}\
  }\textbf {\bibinfo {volume} {87}},\ \bibinfo {pages} {042104} (\bibinfo
  {year} {2013})}\BibitemShut {NoStop}%
\bibitem [{\citenamefont {Ngampruetikorn}\ \emph {et~al.}(2020)\citenamefont
  {Ngampruetikorn}, \citenamefont {Schwab},\ and\ \citenamefont
  {Stephens}}]{Ngampruetikorn-01}%
  \BibitemOpen
  \bibfield  {author} {\bibinfo {author} {\bibfnamefont {Vudtiwat}\
  \bibnamefont {Ngampruetikorn}}, \bibinfo {author} {\bibfnamefont {David~J}\
  \bibnamefont {Schwab}}, \ and\ \bibinfo {author} {\bibfnamefont {Greg~J}\
  \bibnamefont {Stephens}},\ }\bibfield  {title} {\enquote {\bibinfo {title}
  {Energy consumption and cooperation for optimal sensing},}\ }\href@noop {}
  {\bibfield  {journal} {\bibinfo  {journal} {Nat. Commun.}\ }\textbf {\bibinfo
  {volume} {11}},\ \bibinfo {pages} {1--8} (\bibinfo {year}
  {2020})}\BibitemShut {NoStop}%
\bibitem [{\citenamefont {Still}(2020)}]{Still-01}%
  \BibitemOpen
  \bibfield  {author} {\bibinfo {author} {\bibfnamefont {Susanne}\ \bibnamefont
  {Still}},\ }\bibfield  {title} {\enquote {\bibinfo {title} {Thermodynamic
  cost and benefit of memory},}\ }\href@noop {} {\bibfield  {journal} {\bibinfo
   {journal} {Phys. Rev. Lett.}\ }\textbf {\bibinfo {volume} {124}},\ \bibinfo
  {pages} {050601} (\bibinfo {year} {2020})}\BibitemShut {NoStop}%
\bibitem [{\citenamefont {Still}\ \emph {et~al.}(2012)\citenamefont {Still},
  \citenamefont {Sivak}, \citenamefont {Bell},\ and\ \citenamefont
  {Crooks}}]{Still-02}%
  \BibitemOpen
  \bibfield  {author} {\bibinfo {author} {\bibfnamefont {Susanne}\ \bibnamefont
  {Still}}, \bibinfo {author} {\bibfnamefont {D.~A.}\ \bibnamefont {Sivak}},
  \bibinfo {author} {\bibfnamefont {A.~J.}\ \bibnamefont {Bell}}, \ and\
  \bibinfo {author} {\bibfnamefont {G.~E.}\ \bibnamefont {Crooks}},\ }\bibfield
   {title} {\enquote {\bibinfo {title} {Thermodynamics of prediction},}\
  }\href@noop {} {\bibfield  {journal} {\bibinfo  {journal} {Phys. Rev. Lett.}\
  }\textbf {\bibinfo {volume} {109}},\ \bibinfo {pages} {120604} (\bibinfo
  {year} {2012})}\BibitemShut {NoStop}%
\bibitem [{\citenamefont {Marzen}\ and\ \citenamefont
  {Crutchfield}(2018)}]{Marzen-01}%
  \BibitemOpen
  \bibfield  {author} {\bibinfo {author} {\bibfnamefont {S.~E.}\ \bibnamefont
  {Marzen}}\ and\ \bibinfo {author} {\bibfnamefont {J.~P.}\ \bibnamefont
  {Crutchfield}},\ }\bibfield  {title} {\enquote {\bibinfo {title} {Optimized
  bacteria are environmental prediction engines},}\ }\href@noop {} {\bibfield
  {journal} {\bibinfo  {journal} {Phys. Rev. E}\ }\textbf {\bibinfo {volume}
  {98}},\ \bibinfo {pages} {012408} (\bibinfo {year} {2018})}\BibitemShut
  {NoStop}%
\bibitem [{\citenamefont {Schnakenberg}(1976)}]{Schnakenberg-01}%
  \BibitemOpen
  \bibfield  {author} {\bibinfo {author} {\bibfnamefont {J}~\bibnamefont
  {Schnakenberg}},\ }\bibfield  {title} {\enquote {\bibinfo {title} {Network
  theory of microscopic and macroscopic behavior of master equation systems},}\
  }\href@noop {} {\bibfield  {journal} {\bibinfo  {journal} {Rev. Mod. Phys.}\
  }\textbf {\bibinfo {volume} {48}},\ \bibinfo {pages} {571} (\bibinfo {year}
  {1976})}\BibitemShut {NoStop}%
\bibitem [{\citenamefont {Skinner}\ and\ \citenamefont
  {Dunkel}(2021)}]{Skinner-01}%
  \BibitemOpen
  \bibfield  {author} {\bibinfo {author} {\bibfnamefont {DJ}~\bibnamefont
  {Skinner}}\ and\ \bibinfo {author} {\bibfnamefont {J}~\bibnamefont
  {Dunkel}},\ }\bibfield  {title} {\enquote {\bibinfo {title} {Improved bounds
  on entropy production in living systems},}\ }\href@noop {} {\bibfield
  {journal} {\bibinfo  {journal} {Proc. Natl. Acad. Sci.}\ }\textbf {\bibinfo
  {volume} {118}} (\bibinfo {year} {2021})}\BibitemShut {NoStop}%
\bibitem [{\citenamefont {Onsager}(1931)}]{Onsager-01}%
  \BibitemOpen
  \bibfield  {author} {\bibinfo {author} {\bibfnamefont {L}~\bibnamefont
  {Onsager}},\ }\bibfield  {title} {\enquote {\bibinfo {title} {Reciprocal
  relations in irreversible processes. {I}.}}\ }\href@noop {} {\bibfield
  {journal} {\bibinfo  {journal} {Phys. Rev.}\ }\textbf {\bibinfo {volume}
  {37}},\ \bibinfo {pages} {405} (\bibinfo {year} {1931})}\BibitemShut
  {NoStop}%
\bibitem [{\citenamefont {Prigogine}(1945)}]{Prigogine-01}%
  \BibitemOpen
  \bibfield  {author} {\bibinfo {author} {\bibfnamefont {I}~\bibnamefont
  {Prigogine}},\ }\bibfield  {title} {\enquote {\bibinfo {title}
  {Mod{\'e}ration et transformation irr{\'e}versible des syst{\`e}mes
  ouverts},}\ }\href@noop {} {\bibfield  {journal} {\bibinfo  {journal} {Acad.
  R. Belg. Bull. Cl. Sci.}\ }\textbf {\bibinfo {volume} {31}},\ \bibinfo
  {pages} {600--606} (\bibinfo {year} {1945})}\BibitemShut {NoStop}%
\bibitem [{\citenamefont {Strong}\ \emph {et~al.}(1998)\citenamefont {Strong},
  \citenamefont {Koberle}, \citenamefont {de~Ruyter~van Steveninck},\ and\
  \citenamefont {Bialek}}]{Strong-01}%
  \BibitemOpen
  \bibfield  {author} {\bibinfo {author} {\bibfnamefont {S~P}\ \bibnamefont
  {Strong}}, \bibinfo {author} {\bibfnamefont {R}~\bibnamefont {Koberle}},
  \bibinfo {author} {\bibfnamefont {R~R}\ \bibnamefont {de~Ruyter~van
  Steveninck}}, \ and\ \bibinfo {author} {\bibfnamefont {W}~\bibnamefont
  {Bialek}},\ }\bibfield  {title} {\enquote {\bibinfo {title} {Entropy and
  information in neural spike trains},}\ }\href@noop {} {\bibfield  {journal}
  {\bibinfo  {journal} {Phys. Rev. Lett.}\ }\textbf {\bibinfo {volume} {80}},\
  \bibinfo {pages} {197} (\bibinfo {year} {1998})}\BibitemShut {NoStop}%
\bibitem [{\citenamefont {Schneidman}\ \emph {et~al.}(2006)\citenamefont
  {Schneidman}, \citenamefont {Berry~II}, \citenamefont {Segev},\ and\
  \citenamefont {Bialek}}]{Schneidman-01}%
  \BibitemOpen
  \bibfield  {author} {\bibinfo {author} {\bibfnamefont {E}~\bibnamefont
  {Schneidman}}, \bibinfo {author} {\bibfnamefont {M.~J.}\ \bibnamefont
  {Berry~II}}, \bibinfo {author} {\bibfnamefont {R}~\bibnamefont {Segev}}, \
  and\ \bibinfo {author} {\bibfnamefont {W}~\bibnamefont {Bialek}},\ }\bibfield
   {title} {\enquote {\bibinfo {title} {Weak pairwise correlations imply
  strongly correlated network states in a neural population},}\ }\href@noop {}
  {\bibfield  {journal} {\bibinfo  {journal} {Nature}\ }\textbf {\bibinfo
  {volume} {440}},\ \bibinfo {pages} {1007--1012} (\bibinfo {year}
  {2006})}\BibitemShut {NoStop}%
\bibitem [{\citenamefont {Battle}\ \emph {et~al.}(2016)\citenamefont {Battle},
  \citenamefont {Broedersz}, \citenamefont {Fakhri}, \citenamefont {Geyer},
  \citenamefont {Howard}, \citenamefont {Schmidt},\ and\ \citenamefont
  {MacKintosh}}]{Battle-01}%
  \BibitemOpen
  \bibfield  {author} {\bibinfo {author} {\bibfnamefont {C}~\bibnamefont
  {Battle}}, \bibinfo {author} {\bibfnamefont {CP}~\bibnamefont {Broedersz}},
  \bibinfo {author} {\bibfnamefont {N}~\bibnamefont {Fakhri}}, \bibinfo
  {author} {\bibfnamefont {VF}~\bibnamefont {Geyer}}, \bibinfo {author}
  {\bibfnamefont {J}~\bibnamefont {Howard}}, \bibinfo {author} {\bibfnamefont
  {CF}~\bibnamefont {Schmidt}}, \ and\ \bibinfo {author} {\bibfnamefont
  {FC}~\bibnamefont {MacKintosh}},\ }\bibfield  {title} {\enquote {\bibinfo
  {title} {Broken detailed balance at mesoscopic scales in active biological
  systems},}\ }\href@noop {} {\bibfield  {journal} {\bibinfo  {journal}
  {Science}\ }\textbf {\bibinfo {volume} {352}},\ \bibinfo {pages} {604--607}
  (\bibinfo {year} {2016})}\BibitemShut {NoStop}%
\bibitem [{\citenamefont {Lynn}\ \emph {et~al.}(2021)\citenamefont {Lynn},
  \citenamefont {Cornblath}, \citenamefont {Papadopoulos}, \citenamefont
  {Bertolero},\ and\ \citenamefont {Bassett}}]{Lynn-09}%
  \BibitemOpen
  \bibfield  {author} {\bibinfo {author} {\bibfnamefont {Christopher~W}\
  \bibnamefont {Lynn}}, \bibinfo {author} {\bibfnamefont {E~J}\ \bibnamefont
  {Cornblath}}, \bibinfo {author} {\bibfnamefont {L}~\bibnamefont
  {Papadopoulos}}, \bibinfo {author} {\bibfnamefont {M~A}\ \bibnamefont
  {Bertolero}}, \ and\ \bibinfo {author} {\bibfnamefont {Danielle~S}\
  \bibnamefont {Bassett}},\ }\bibfield  {title} {\enquote {\bibinfo {title}
  {Broken detailed balance and entropy production in the human brain},}\
  }\href@noop {} {\bibfield  {journal} {\bibinfo  {journal} {Proc. Natl. Acad.
  Sci.}\ }\textbf {\bibinfo {volume} {118}} (\bibinfo {year}
  {2021})}\BibitemShut {NoStop}%
\bibitem [{\citenamefont {Mart{\'\i}nez}\ \emph {et~al.}(2019)\citenamefont
  {Mart{\'\i}nez}, \citenamefont {Bisker}, \citenamefont {Horowitz},\ and\
  \citenamefont {Parrondo}}]{Martinez-01}%
  \BibitemOpen
  \bibfield  {author} {\bibinfo {author} {\bibfnamefont {Ignacio~A}\
  \bibnamefont {Mart{\'\i}nez}}, \bibinfo {author} {\bibfnamefont {Gili}\
  \bibnamefont {Bisker}}, \bibinfo {author} {\bibfnamefont {Jordan~M}\
  \bibnamefont {Horowitz}}, \ and\ \bibinfo {author} {\bibfnamefont {Juan~MR}\
  \bibnamefont {Parrondo}},\ }\bibfield  {title} {\enquote {\bibinfo {title}
  {Inferring broken detailed balance in the absence of observable currents},}\
  }\href@noop {} {\bibfield  {journal} {\bibinfo  {journal} {Nat. Commun.}\
  }\textbf {\bibinfo {volume} {10}},\ \bibinfo {pages} {1--10} (\bibinfo {year}
  {2019})}\BibitemShut {NoStop}%
\bibitem [{\citenamefont {Li}\ \emph {et~al.}(2019)\citenamefont {Li},
  \citenamefont {Horowitz}, \citenamefont {Gingrich},\ and\ \citenamefont
  {Fakhri}}]{Li-01}%
  \BibitemOpen
  \bibfield  {author} {\bibinfo {author} {\bibfnamefont {Junang}\ \bibnamefont
  {Li}}, \bibinfo {author} {\bibfnamefont {Jordan~M}\ \bibnamefont {Horowitz}},
  \bibinfo {author} {\bibfnamefont {Todd~R}\ \bibnamefont {Gingrich}}, \ and\
  \bibinfo {author} {\bibfnamefont {Nikta}\ \bibnamefont {Fakhri}},\ }\bibfield
   {title} {\enquote {\bibinfo {title} {Quantifying dissipation using
  fluctuating currents},}\ }\href@noop {} {\bibfield  {journal} {\bibinfo
  {journal} {Nat. Commun.}\ }\textbf {\bibinfo {volume} {10}},\ \bibinfo
  {pages} {1--9} (\bibinfo {year} {2019})}\BibitemShut {NoStop}%
\bibitem [{\citenamefont {Bialek}\ \emph {et~al.}(2012)\citenamefont {Bialek},
  \citenamefont {Cavagna}, \citenamefont {Giardina}, \citenamefont {Mora},
  \citenamefont {Silvestri}, \citenamefont {Viale},\ and\ \citenamefont
  {Walczak}}]{Bialek-02}%
  \BibitemOpen
  \bibfield  {author} {\bibinfo {author} {\bibfnamefont {W}~\bibnamefont
  {Bialek}}, \bibinfo {author} {\bibfnamefont {A}~\bibnamefont {Cavagna}},
  \bibinfo {author} {\bibfnamefont {I}~\bibnamefont {Giardina}}, \bibinfo
  {author} {\bibfnamefont {T}~\bibnamefont {Mora}}, \bibinfo {author}
  {\bibfnamefont {E}~\bibnamefont {Silvestri}}, \bibinfo {author}
  {\bibfnamefont {M}~\bibnamefont {Viale}}, \ and\ \bibinfo {author}
  {\bibfnamefont {A~M}\ \bibnamefont {Walczak}},\ }\bibfield  {title} {\enquote
  {\bibinfo {title} {Statistical mechanics for natural flocks of birds},}\
  }\href@noop {} {\bibfield  {journal} {\bibinfo  {journal} {Proc. Natl. Acd.
  Sci.}\ }\textbf {\bibinfo {volume} {109}},\ \bibinfo {pages} {4786--4791}
  (\bibinfo {year} {2012})}\BibitemShut {NoStop}%
\bibitem [{\citenamefont {Meshulam}\ \emph {et~al.}(2017)\citenamefont
  {Meshulam}, \citenamefont {Gauthier}, \citenamefont {Brody}, \citenamefont
  {Tank},\ and\ \citenamefont {Bialek}}]{Meshulam-17}%
  \BibitemOpen
  \bibfield  {author} {\bibinfo {author} {\bibfnamefont {L}~\bibnamefont
  {Meshulam}}, \bibinfo {author} {\bibfnamefont {J~L}\ \bibnamefont
  {Gauthier}}, \bibinfo {author} {\bibfnamefont {C~D}\ \bibnamefont {Brody}},
  \bibinfo {author} {\bibfnamefont {D~W}\ \bibnamefont {Tank}}, \ and\ \bibinfo
  {author} {\bibfnamefont {W}~\bibnamefont {Bialek}},\ }\bibfield  {title}
  {\enquote {\bibinfo {title} {Collective behavior of place and non-place
  neurons in the hippocampal network},}\ }\href@noop {} {\bibfield  {journal}
  {\bibinfo  {journal} {Neuron}\ }\textbf {\bibinfo {volume} {96}},\ \bibinfo
  {pages} {1178--1191} (\bibinfo {year} {2017})}\BibitemShut {NoStop}%
\bibitem [{\citenamefont {Lynn}\ \emph {et~al.}(2019)\citenamefont {Lynn},
  \citenamefont {Papadopoulos}, \citenamefont {Lee},\ and\ \citenamefont
  {Bassett}}]{Lynn-04}%
  \BibitemOpen
  \bibfield  {author} {\bibinfo {author} {\bibfnamefont {Christopher~W}\
  \bibnamefont {Lynn}}, \bibinfo {author} {\bibfnamefont {Lia}\ \bibnamefont
  {Papadopoulos}}, \bibinfo {author} {\bibfnamefont {Daniel~D}\ \bibnamefont
  {Lee}}, \ and\ \bibinfo {author} {\bibfnamefont {Danielle~S}\ \bibnamefont
  {Bassett}},\ }\bibfield  {title} {\enquote {\bibinfo {title} {Surges of
  collective human activity emerge from simple pairwise correlations},}\
  }\href@noop {} {\bibfield  {journal} {\bibinfo  {journal} {Phys. Rev. X}\
  }\textbf {\bibinfo {volume} {9}},\ \bibinfo {pages} {011022} (\bibinfo {year}
  {2019})}\BibitemShut {NoStop}%
\bibitem [{\citenamefont {Stuhrmann}\ \emph {et~al.}(2012)\citenamefont
  {Stuhrmann}, \citenamefont {e~Silva}, \citenamefont {Depken}, \citenamefont
  {MacKintosh},\ and\ \citenamefont {Koenderink}}]{Stuhrmann-01}%
  \BibitemOpen
  \bibfield  {author} {\bibinfo {author} {\bibfnamefont {Bj{\"o}rn}\
  \bibnamefont {Stuhrmann}}, \bibinfo {author} {\bibfnamefont {Marina~Soares}\
  \bibnamefont {e~Silva}}, \bibinfo {author} {\bibfnamefont {Martin}\
  \bibnamefont {Depken}}, \bibinfo {author} {\bibfnamefont {Fred~C}\
  \bibnamefont {MacKintosh}}, \ and\ \bibinfo {author} {\bibfnamefont
  {Gijsje~H}\ \bibnamefont {Koenderink}},\ }\bibfield  {title} {\enquote
  {\bibinfo {title} {Nonequilibrium fluctuations of a remodeling in vitro
  cytoskeleton},}\ }\href@noop {} {\bibfield  {journal} {\bibinfo  {journal}
  {Phys. Rev. E}\ }\textbf {\bibinfo {volume} {86}},\ \bibinfo {pages}
  {020901(R)} (\bibinfo {year} {2012})}\BibitemShut {NoStop}%
\bibitem [{\citenamefont {Soares~e Silva}\ \emph {et~al.}(2011)\citenamefont
  {Soares~e Silva}, \citenamefont {Depken}, \citenamefont {Stuhrmann},
  \citenamefont {Korsten}, \citenamefont {MacKintosh},\ and\ \citenamefont
  {Koenderink}}]{Soares-01}%
  \BibitemOpen
  \bibfield  {author} {\bibinfo {author} {\bibfnamefont {Marina~Soares}\
  \bibnamefont {Soares~e Silva}}, \bibinfo {author} {\bibfnamefont {Martin}\
  \bibnamefont {Depken}}, \bibinfo {author} {\bibfnamefont {Bj{\"o}rn}\
  \bibnamefont {Stuhrmann}}, \bibinfo {author} {\bibfnamefont {Marijn}\
  \bibnamefont {Korsten}}, \bibinfo {author} {\bibfnamefont {Fred~C}\
  \bibnamefont {MacKintosh}}, \ and\ \bibinfo {author} {\bibfnamefont
  {Gijsje~H}\ \bibnamefont {Koenderink}},\ }\bibfield  {title} {\enquote
  {\bibinfo {title} {Active multistage coarsening of actin networks driven by
  myosin motors},}\ }\href@noop {} {\bibfield  {journal} {\bibinfo  {journal}
  {Proc. Natl. Acad. Sci.}\ }\textbf {\bibinfo {volume} {108}},\ \bibinfo
  {pages} {9408--9413} (\bibinfo {year} {2011})}\BibitemShut {NoStop}%
\bibitem [{\citenamefont {Fakhri}\ \emph {et~al.}(2014)\citenamefont {Fakhri},
  \citenamefont {Wessel}, \citenamefont {Willms}, \citenamefont {Pasquali},
  \citenamefont {Klopfenstein}, \citenamefont {MacKintosh},\ and\ \citenamefont
  {Schmidt}}]{Fakhri-01}%
  \BibitemOpen
  \bibfield  {author} {\bibinfo {author} {\bibfnamefont {Nikta}\ \bibnamefont
  {Fakhri}}, \bibinfo {author} {\bibfnamefont {Alok~D}\ \bibnamefont {Wessel}},
  \bibinfo {author} {\bibfnamefont {Charlotte}\ \bibnamefont {Willms}},
  \bibinfo {author} {\bibfnamefont {Matteo}\ \bibnamefont {Pasquali}}, \bibinfo
  {author} {\bibfnamefont {Dieter~R}\ \bibnamefont {Klopfenstein}}, \bibinfo
  {author} {\bibfnamefont {Frederick~C}\ \bibnamefont {MacKintosh}}, \ and\
  \bibinfo {author} {\bibfnamefont {Christoph~F}\ \bibnamefont {Schmidt}},\
  }\bibfield  {title} {\enquote {\bibinfo {title} {High-resolution mapping of
  intracellular fluctuations using carbon nanotubes},}\ }\href@noop {}
  {\bibfield  {journal} {\bibinfo  {journal} {Science}\ }\textbf {\bibinfo
  {volume} {344}},\ \bibinfo {pages} {1031--1035} (\bibinfo {year}
  {2014})}\BibitemShut {NoStop}%
\bibitem [{\citenamefont {Lynn}\ and\ \citenamefont {Bassett}(2019)}]{Lynn-05}%
  \BibitemOpen
  \bibfield  {author} {\bibinfo {author} {\bibfnamefont {Christopher~W}\
  \bibnamefont {Lynn}}\ and\ \bibinfo {author} {\bibfnamefont {Danielle~S}\
  \bibnamefont {Bassett}},\ }\bibfield  {title} {\enquote {\bibinfo {title}
  {The physics of brain network structure, function and control},}\ }\href@noop
  {} {\bibfield  {journal} {\bibinfo  {journal} {Nat. Rev. Phys.}\ }\textbf
  {\bibinfo {volume} {1}},\ \bibinfo {pages} {318} (\bibinfo {year}
  {2019})}\BibitemShut {NoStop}%
\bibitem [{\citenamefont {Cavagna}\ \emph {et~al.}(2014)\citenamefont
  {Cavagna}, \citenamefont {Giardina}, \citenamefont {Ginelli}, \citenamefont
  {Mora}, \citenamefont {Piovani}, \citenamefont {Tavarone},\ and\
  \citenamefont {Walczak}}]{Cavagna-01}%
  \BibitemOpen
  \bibfield  {author} {\bibinfo {author} {\bibfnamefont {A}~\bibnamefont
  {Cavagna}}, \bibinfo {author} {\bibfnamefont {I}~\bibnamefont {Giardina}},
  \bibinfo {author} {\bibfnamefont {F}~\bibnamefont {Ginelli}}, \bibinfo
  {author} {\bibfnamefont {T}~\bibnamefont {Mora}}, \bibinfo {author}
  {\bibfnamefont {D}~\bibnamefont {Piovani}}, \bibinfo {author} {\bibfnamefont
  {R}~\bibnamefont {Tavarone}}, \ and\ \bibinfo {author} {\bibfnamefont
  {A.~M.}\ \bibnamefont {Walczak}},\ }\bibfield  {title} {\enquote {\bibinfo
  {title} {Dynamical maximum entropy approach to flocking},}\ }\href@noop {}
  {\bibfield  {journal} {\bibinfo  {journal} {Phys. Rev. E}\ }\textbf {\bibinfo
  {volume} {89}},\ \bibinfo {pages} {042707} (\bibinfo {year}
  {2014})}\BibitemShut {NoStop}%
\bibitem [{\citenamefont {Mitchell}\ \emph {et~al.}(2013)\citenamefont
  {Mitchell}, \citenamefont {Lange},\ and\ \citenamefont {Brus}}]{Mitchell-01}%
  \BibitemOpen
  \bibfield  {author} {\bibinfo {author} {\bibfnamefont {Sara~McLaughlin}\
  \bibnamefont {Mitchell}}, \bibinfo {author} {\bibfnamefont {Samantha}\
  \bibnamefont {Lange}}, \ and\ \bibinfo {author} {\bibfnamefont {Holly}\
  \bibnamefont {Brus}},\ }\bibfield  {title} {\enquote {\bibinfo {title}
  {Gendered citation patterns in international relations journals},}\
  }\href@noop {} {\bibfield  {journal} {\bibinfo  {journal} {Int. Stud.
  Perspect.}\ }\textbf {\bibinfo {volume} {14}},\ \bibinfo {pages} {485--492}
  (\bibinfo {year} {2013})}\BibitemShut {NoStop}%
\bibitem [{\citenamefont {Dion}\ \emph {et~al.}(2018)\citenamefont {Dion},
  \citenamefont {Sumner},\ and\ \citenamefont {Mitchell}}]{Dion-01}%
  \BibitemOpen
  \bibfield  {author} {\bibinfo {author} {\bibfnamefont {Michelle~L}\
  \bibnamefont {Dion}}, \bibinfo {author} {\bibfnamefont {Jane~Lawrence}\
  \bibnamefont {Sumner}}, \ and\ \bibinfo {author} {\bibfnamefont
  {Sara~McLaughlin}\ \bibnamefont {Mitchell}},\ }\bibfield  {title} {\enquote
  {\bibinfo {title} {Gendered citation patterns across political science and
  social science methodology fields},}\ }\href@noop {} {\bibfield  {journal}
  {\bibinfo  {journal} {Polit. Anal.}\ }\textbf {\bibinfo {volume} {26}},\
  \bibinfo {pages} {312--327} (\bibinfo {year} {2018})}\BibitemShut {NoStop}%
\bibitem [{\citenamefont {Caplar}\ \emph {et~al.}(2017)\citenamefont {Caplar},
  \citenamefont {Tacchella},\ and\ \citenamefont {Birrer}}]{Caplar-01}%
  \BibitemOpen
  \bibfield  {author} {\bibinfo {author} {\bibfnamefont {Neven}\ \bibnamefont
  {Caplar}}, \bibinfo {author} {\bibfnamefont {Sandro}\ \bibnamefont
  {Tacchella}}, \ and\ \bibinfo {author} {\bibfnamefont {Simon}\ \bibnamefont
  {Birrer}},\ }\bibfield  {title} {\enquote {\bibinfo {title} {Quantitative
  evaluation of gender bias in astronomical publications from citation
  counts},}\ }\href@noop {} {\bibfield  {journal} {\bibinfo  {journal} {Nat.
  Astron.}\ }\textbf {\bibinfo {volume} {1}},\ \bibinfo {pages} {1--5}
  (\bibinfo {year} {2017})}\BibitemShut {NoStop}%
\bibitem [{\citenamefont {Dworkin}\ \emph {et~al.}(2020)\citenamefont
  {Dworkin}, \citenamefont {Linn}, \citenamefont {Teich}, \citenamefont {Zurn},
  \citenamefont {Shinohara},\ and\ \citenamefont {Bassett}}]{Dworkin-01}%
  \BibitemOpen
  \bibfield  {author} {\bibinfo {author} {\bibfnamefont {Jordan~D}\
  \bibnamefont {Dworkin}}, \bibinfo {author} {\bibfnamefont {Kristin~A}\
  \bibnamefont {Linn}}, \bibinfo {author} {\bibfnamefont {Erin~G}\ \bibnamefont
  {Teich}}, \bibinfo {author} {\bibfnamefont {Perry}\ \bibnamefont {Zurn}},
  \bibinfo {author} {\bibfnamefont {Russell~T}\ \bibnamefont {Shinohara}}, \
  and\ \bibinfo {author} {\bibfnamefont {Danielle~S}\ \bibnamefont {Bassett}},\
  }\bibfield  {title} {\enquote {\bibinfo {title} {The extent and drivers of
  gender imbalance in neuroscience reference lists},}\ }\href@noop {}
  {\bibfield  {journal} {\bibinfo  {journal} {Nat Neurosci.}\ }\textbf
  {\bibinfo {volume} {23}},\ \bibinfo {pages} {918--926} (\bibinfo {year}
  {2020})}\BibitemShut {NoStop}%
\bibitem [{\citenamefont {Bertolero}\ \emph {et~al.}(2020)\citenamefont
  {Bertolero}, \citenamefont {Dworkin}, \citenamefont {David}, \citenamefont
  {Lloreda}, \citenamefont {Srivastava}, \citenamefont {Stiso}, \citenamefont
  {Zhou}, \citenamefont {Dzirasa}, \citenamefont {Fair}, \citenamefont
  {Kaczkurkin} \emph {et~al.}}]{Bertolero-01}%
  \BibitemOpen
  \bibfield  {author} {\bibinfo {author} {\bibfnamefont {Maxwell~A}\
  \bibnamefont {Bertolero}}, \bibinfo {author} {\bibfnamefont {Jordan~D}\
  \bibnamefont {Dworkin}}, \bibinfo {author} {\bibfnamefont {Sophia~U}\
  \bibnamefont {David}}, \bibinfo {author} {\bibfnamefont {Claudia~L{\'o}pez}\
  \bibnamefont {Lloreda}}, \bibinfo {author} {\bibfnamefont {Pragya}\
  \bibnamefont {Srivastava}}, \bibinfo {author} {\bibfnamefont {Jennifer}\
  \bibnamefont {Stiso}}, \bibinfo {author} {\bibfnamefont {Dale}\ \bibnamefont
  {Zhou}}, \bibinfo {author} {\bibfnamefont {Kafui}\ \bibnamefont {Dzirasa}},
  \bibinfo {author} {\bibfnamefont {Damien~A}\ \bibnamefont {Fair}}, \bibinfo
  {author} {\bibfnamefont {Antonia~N}\ \bibnamefont {Kaczkurkin}},  \emph
  {et~al.},\ }\bibfield  {title} {\enquote {\bibinfo {title} {Racial and ethnic
  imbalance in neuroscience reference lists and intersections with gender},}\
  }\href@noop {} {\bibfield  {journal} {\bibinfo  {journal} {bioRxiv}\ }
  (\bibinfo {year} {2020})}\BibitemShut {NoStop}%
\bibitem [{\citenamefont {Teich}\ \emph {et~al.}(2021)\citenamefont {Teich},
  \citenamefont {Kim}, \citenamefont {Lynn}, \citenamefont {Simon},
  \citenamefont {Klishin}, \citenamefont {Szymula}, \citenamefont {Srivastava},
  \citenamefont {Bassett}, \citenamefont {Zurn}, \citenamefont {Dworkin} \emph
  {et~al.}}]{Teich-01}%
  \BibitemOpen
  \bibfield  {author} {\bibinfo {author} {\bibfnamefont {Erin~G}\ \bibnamefont
  {Teich}}, \bibinfo {author} {\bibfnamefont {Jason~Z}\ \bibnamefont {Kim}},
  \bibinfo {author} {\bibfnamefont {Christopher~W}\ \bibnamefont {Lynn}},
  \bibinfo {author} {\bibfnamefont {Samantha~C}\ \bibnamefont {Simon}},
  \bibinfo {author} {\bibfnamefont {Andrei~A}\ \bibnamefont {Klishin}},
  \bibinfo {author} {\bibfnamefont {Karol~P}\ \bibnamefont {Szymula}}, \bibinfo
  {author} {\bibfnamefont {Pragya}\ \bibnamefont {Srivastava}}, \bibinfo
  {author} {\bibfnamefont {Lee~C}\ \bibnamefont {Bassett}}, \bibinfo {author}
  {\bibfnamefont {Perry}\ \bibnamefont {Zurn}}, \bibinfo {author}
  {\bibfnamefont {Jordan~D}\ \bibnamefont {Dworkin}},  \emph {et~al.},\
  }\bibfield  {title} {\enquote {\bibinfo {title} {Citation inequity and
  gendered citation practices in contemporary physics},}\ }\href@noop {}
  {\bibfield  {journal} {\bibinfo  {journal} {arXiv preprint arXiv:2112.09047}\
  } (\bibinfo {year} {2021})}\BibitemShut {NoStop}%
\bibitem [{\citenamefont {Rold{\'a}n}\ and\ \citenamefont
  {Parrondo}(2010)}]{Roldan-01}%
  \BibitemOpen
  \bibfield  {author} {\bibinfo {author} {\bibfnamefont {{\'E}dgar}\
  \bibnamefont {Rold{\'a}n}}\ and\ \bibinfo {author} {\bibfnamefont {Juan~MR}\
  \bibnamefont {Parrondo}},\ }\bibfield  {title} {\enquote {\bibinfo {title}
  {Estimating dissipation from single stationary trajectories},}\ }\href@noop
  {} {\bibfield  {journal} {\bibinfo  {journal} {Phys. Rev. Lett.}\ }\textbf
  {\bibinfo {volume} {105}},\ \bibinfo {pages} {150607} (\bibinfo {year}
  {2010})}\BibitemShut {NoStop}%
\end{thebibliography}%

\end{document}